\documentclass[useAmain sequence,usenatbib]{mn2e}
\usepackage[colorlinks=true,citecolor=blue]{hyperref}
\usepackage{newtxtext,newtxmath}
\usepackage{ifpdf}
\ifpdf
 \usepackage[pdftex]{graphicx}
 \usepackage{epstopdf}
 \AppendGraphicsExtensions{.ps}
\else
 \usepackage{graphicx}
\fi

\usepackage{subfigure}
\usepackage{float}
\usepackage{deluxetable}
\usepackage{longtable}

\DeclareMathAlphabet{\mathsc}{OT1}{cmr}{m}{sc}
\def\testbx{bx}%
\DeclareRobustCommand{\ion}[2]{%
	\relax\ifmmode
	\ifx\testbx\f@series
	{\mathbf{#1\,\mathsc{#2}}}\else
	{\mathrm{#1\,\mathsc{#2}}}\fi
	\else\textup{#1\,{\mdseries\textsc{#2}}}%
	\fi}
\newcommand{\Hi} {\ion{H}{i}}

\newcommand{\ha} {\mbox{H$\alpha$}}
\newcommand{\hb} {\mbox{H$\beta$}}

\newcommand{\Hei} {\ion{He}{i}}
\newcommand{\Heii} {\ion{He}{ii}}

\newcommand{\Nii} {\ion{N}{ii}}

\newcommand{\Oi} {\ion{O}{i}}

\newcommand{\Nai} {\ion{Na}{i}}

\newcommand{\Caii} {\ion{Ca}{ii}}
\newcommand{\Scii} {\ion{Sc}{ii}}
\newcommand{\Tiii} {\ion{Ti}{ii}}

\newcommand{\Feii} {\ion{Fe}{ii}}

\newcommand{\Baii} {\ion{Ba}{ii}}

\newcommand{\sn}{ASASSN-18am}
\newcommand{\host}{WISE~J163554.27+400151.8}
\newcommand{\EpEpoch}{JD~2,458,130.6}
\newcommand{\lc}{light-curve}

\newcommand{\synow}{\textsc{synow}}

\newcommand{\iraf}{\texttt{IRAF}}
\newcommand{\daophot}{\textsc{daophot}}

\newcommand{\ebv}{\mbox{$E(B-V)$}}

\newcommand{\maghundred}{\mbox{mag (100 d)$ ^{-1} $}}
\newcommand{\msun}{\mbox{M$_{\odot}$}}

\newcommand{\kms}{\mbox{$\rm{\,km\,s^{-1}}$}}

\newcommand{\ergs}{\mbox{$\rm{\,erg\,s^{-1}}$}}

\newcommand{\nickel}{\mbox{$^{56}$Ni}}
\newcommand{\cobalt}{\mbox{$^{56}$Co}}
\newcommand{\iron}{\mbox{$^{56}$Fe}}

\newcommand{\ld}{\mbox{$\lambda$}}
\newcommand{\ldld}{\mbox{$\lambda\lambda$}}
\newcommand{\swift}{{\it Swift}}

\begin{document}

\title[Luminous Type~II supernova \sn]
{\sn/SN~2018gk: An overluminous Type IIb supernova from a massive progenitor}
\author[Bose et al.]
{Subhash Bose,$^{1, 2}$\thanks{e-mail: email@subhashbose.com, bose.48@osu.edu}
	Subo Dong,$^{3}$\thanks{e-mail: dongsubo@pku.edu.cn}
	C. S. Kochanek,$^{1, 2}$
	M. D. Stritzinger,$^{4}$
	Chris Ashall,$^{5}$
	\newauthor
	Stefano Benetti,$^{6}$
	E. Falco,$^{7}$
	Alexei V. Filippenko,$^{8, 9}$
	Andrea Pastorello,$^{6}$
	\newauthor
	Jose L. Prieto,$^{10, 11}$
	Auni Somero,$^{12}$
	Tuguldur Sukhbold,$^{1, 2}$
	Junbo Zhang,$^{13}$
	\newauthor
	Katie Auchettl,$^{14, 15, 16, 17}$
	Thomas G. Brink,$^{8}$
	J. S. Brown,$^{18}$
	Ping Chen,$^{3}$
	A. Fiore,$^{6, 19}$
	\newauthor
	Dirk Grupe,$^{20}$
	T. W.-S. Holoien,$^{21}$
	Peter Lundqvist,$^{22}$
	Seppo Mattila,$^{12}$
	\newauthor
	Robert Mutel,$^{23}$
	David Pooley,$^{24}$
	R. S. Post,$^{25}$
        Naveen Reddy,$^{26}$
	Thomas M. Reynolds,$^{12}$
	\newauthor
	Benjamin J. Shappee,$^{27}$
	K. Z. Stanek,$^{1, 2}$
	Todd A. Thompson,$^{1, 2}$
	S. Villanueva Jr.,$^{1}$
	\newauthor
	and WeiKang Zheng$^{8}$
	\\ \\
	Affiliations are listed at the end of the paper
}

\date{Accepted.....; Received .....}

\pagerange{\pageref{firstpage}--\pageref{lastpage}} \pubyear{}

\maketitle

\label{firstpage}

\pagebreak

\begin{abstract}
\sn/SN~2018gk is a newly discovered member of the rare group of luminous, hydrogen-rich supernovae (SNe) with a peak absolute magnitude of $M_V \approx -20$\,mag that is in between normal core-collapse SNe and superluminous SNe. These SNe show no prominent spectroscopic signatures of ejecta interacting with circumstellar material (CSM), and their powering mechanism is debated. \sn\ declines extremely rapidly for a Type~II SN, with a photospheric-phase decline rate of $\sim6.0$\,\maghundred.
Owing to the weakening of \Hi\ and the appearance of \Hei\ in its later phases, \sn\ is spectroscopically a Type~IIb SN with a partially stripped envelope. 
However, its photometric and spectroscopic evolution show significant differences from typical SNe IIb. Using a radiative diffusion model, we find that the light curve requires a high synthesised \nickel\ mass $M_{\rm Ni} \sim0.4\, \msun$ and ejecta with high kinetic energy $E_{\rm kin} =$ (7--10) $\times10^{51} $\,erg. Introducing a magnetar central engine still requires $M_{\rm Ni} \sim0.3\, \msun$ and $E_{\rm kin}= 3\times10^{51} $\,erg. The high \nickel\ mass is consistent with strong iron-group nebular lines in its spectra, which are also similar to several SNe~Ic-BL with high \nickel\ yields. The earliest spectrum shows ``flash ionisation" features, from which we estimate a mass-loss rate of $ \dot{M}\approx 2\times10^{-4} \, \rm \msun~yr^{-1} $. This wind density is too low to power the luminous light curve by ejecta-CSM interaction. 
We measure expansion velocities as high as $ 17,000 $\,\kms\ for \ha, which is remarkably high compared to other SNe~II. We estimate an oxygen core mass of 1.8--3.4\,\msun\ using the [\Oi] luminosity measured from a nebular-phase spectrum,
implying a progenitor with a zero-age main sequence mass of 19--26\,\msun. 

\end{abstract}	


\section{Introduction} \label{sec:intro}
Supernovae (SNe) originating from massive stars ($ \gtrsim10\,\msun $) that have retained a significant amount of hydrogen at the time of explosion show strong Balmer lines in their spectra and are classified as Type~II SNe \citep[e.g.,][]{1997ARA&A..35..309F}. Several subclasses have been introduced to this hydrogen-rich class of SNe based on photometric or spectroscopic properties.
Historically, the Type IIP and IIL subclasses \citep{1979A&A....72..287B,1985AJ.....90.2303D} were mainly motivated by light-curve shapes in the photospheric phase ($ \lesssim 100 $\,d), where the former show a distinct ``plateau'' in the \lc\ and the latter show a ``linear'' decline in magnitude. 
With increasing numbers of SNe discovered by systematic surveys, it became increasingly clear that normal SN~II light-curve shapes form a continuous distribution \citep[e.g.,][]{2014ApJ...786...67A, 2014MNRAS.445..554F,2015ApJ...799..208S}, suggesting a continuum in their ejecta properties.
Hereafter, we refer to this entire class as a combined SN~IIP/L class, or with the general designation SNe~II, while we occasionally mention SNe~IIP and IIL as two extremes of the light-curve slope distribution. 

Two additional subclasses of SNe~II, which are differentiated by their spectroscopic properties, are SNe~IIn and IIb. The spectra of SNe~IIn show relatively narrow ($< 100\,\kms$) or intermediate-width ($\sim 1000\,\kms$) emission lines, which are believed to result from strong interactions between the ejecta and circumstellar material (CSM) \citep{1990MNRAS.244..269S}. These interactions act as an additional power source and can produce light curves significantly different from those of normal SNe~II. SNe~IIb constitute a transition class of objects linking SNe~II and Ib \citep{1988AJ.....96.1941F,1993ApJ...415L.103F}. The spectra of SNe IIb show prominent hydrogen lines at early times, while helium lines become visible after a few weeks. The hydrogen envelope of a SN~IIb is thought to be partially stripped so that the helium core is revealed as the envelope becomes optically thin.

The primary powering mechanisms in normal SNe~II are \nickel\ decay and internal energy deposited by the shock in the ejecta. During the post-photospheric phase, when the ejecta become optically thin, light curves of all common SN types (both core-collapse SNe and SNe~Ia) are powered by energy deposition from the radioactive decay chain of $\nickel\rightarrow\cobalt\rightarrow\iron$, with the exception of some SNe that undergo strong ejecta-CSM interactions during late phases. The \lc\ slopes during the radioactive-powered phase depend on the amount of  $ \gamma $-ray leakage, which is determined by the ejecta properties. 

\subsubsection*{Hydrogen-rich luminous SNe (LSNe-II)}

Over the last decade or so, a new subclass of SNe called superluminous supernovae (SLSNe) has emerged, based on their high peak luminosities compared to common SN types. The hydrogen-poor subclass of SLSNe \citep[SLSNe-I;][]{2007ApJ...668L..99Q,2011Natur.474..487Q} have peak luminosities  $M_g \lesssim-20$\,mag \citep{2019ARA&A..57..305G}, which are significantly more luminous than common SNe (SNe~Ia, IIP/L, and Ib/c). There is no systematic study of the luminosity distribution of SLSNe-II, but many known SLSNe-II show strong CSM interaction (SLSNe-IIn), and the average peak luminosities of SLSNe-IIn are estimated to be $\sim -21$\,mag \citep{2019ARA&A..57..305G}.  

In the last few years, a handful of luminous hydrogen-rich SNe have been found with no prominent signatures of CSM interaction and optical luminosities of $\sim -20 $\,mag (e.g., PTF10iam by \citealt{2016ApJ...819...35A}, SN 2013fc by \citealt{2016MNRAS.456..323K}, ASASSN-15nx by \citealt{2018ApJ...862..107B}, and SN 2016gsd by \citealt{2020MNRAS.493.1761R}). Hereafter we refer to these luminous hydrogen-rich SNe as ``LSNe-II''. We stress that it is not yet clear whether LSNe-II form a distinct subclass of SNe~II. \citet{2018ApJ...862..107B} showed that their volumetric rate might be comparable to that of SLSNe based on the single discovery of ASASSN-15nx, providing an intriguing yet tentative possibility that they might be part of a continuous luminosity distribution connecting normal SNe with SLSNe.  

LSNe-II have numerous \lc\ and spectroscopic peculiarities, and it is challenging to explain their luminosities with the commonly proposed mechanisms for core-collapse supernovae (ccSNe). None of these SNe shows persistent narrow emission lines indicating strong ejecta-CSM interaction as seen in luminous SNe~IIn. However, the lack of narrow emission lines may not be sufficient to entirely rule out CSM interaction. In all of the LSN-II studies to date, CSM interaction is considered as at least one of the possible powering sources, where certain CSM configurations may able to hide the strong emission lines and leave only weak
spectroscopic features to be associated with CSM interaction. \cite{2018ApJ...862..107B} also showed that the light curves of ASASSN-15nx could be entirely powered by radioactive decay but required a very large amount of \nickel\ ($ M_{\rm Ni} \approx 1.6 $\,\msun). However, this scenario may not be tenable given the lack of strong lines of iron-group elements in its spectra. Such a high \nickel\ mass is not compatible with the neutrino mechanism of ccSNe \citep[see, e.g.,][]{2016ApJ...821...38S,2020ApJ...890...51E}, but it is possible for collapse-induced thermonuclear explosions \citep{1957RvMP...29..547B,1960ApJ...132..565H,1964ApJS....9..201F,CITE2, CITE3,2015ApJ...811...97K} or pair-instability SNe \citep[e.g.,][]{1967PhRvL..18..379B,2011ApJ...734..102K}. However, pair-instability SN models exhibit extended light curves which are incompatible with ASASSN-15nx and most other LSNe-II.
For ASASSN-15nx, \cite{2019AstL...45..427C} also suggested magnetar spin-down as an alternate powering mechanism based on detailed spectroscopic and light-curve modeling. 

Here, we report the latest addition to this rare and underexplored group of LSNe-II, \sn, with a peak luminosity of $ M_V \approx -19.7 $\,mag.
We present a detailed study of this SN from its discovery to well into the nebular phase.
In \S\ref{sec:dist_ext_host} we discuss the adopted values for the  distance, explosion epoch, and line-of-sight extinction, along with the host-galaxy properties. The data obtained from various telescopes are summarised in \S\ref{sec:data}. The light curve and spectra are analysed in \S\ref{sec:lightcurve} and \S\ref{sec:spectra}, respectively. In \S\ref{sec:lc_discussion}, we discuss various powering mechanisms through modeling. Nebular-phase emission lines are analysed in \S\ref{sec:oxygen_mass}, and we estimate oxygen and zero-age main sequence (ZAMS) masses of the progenitor. We summarise our findings in  \S\ref{sec:summary}.

\section{Explosion epoch, extinction, and host properties} \label{sec:dist_ext_host}

\begin{figure}
	\centering
	\includegraphics[width=\linewidth]{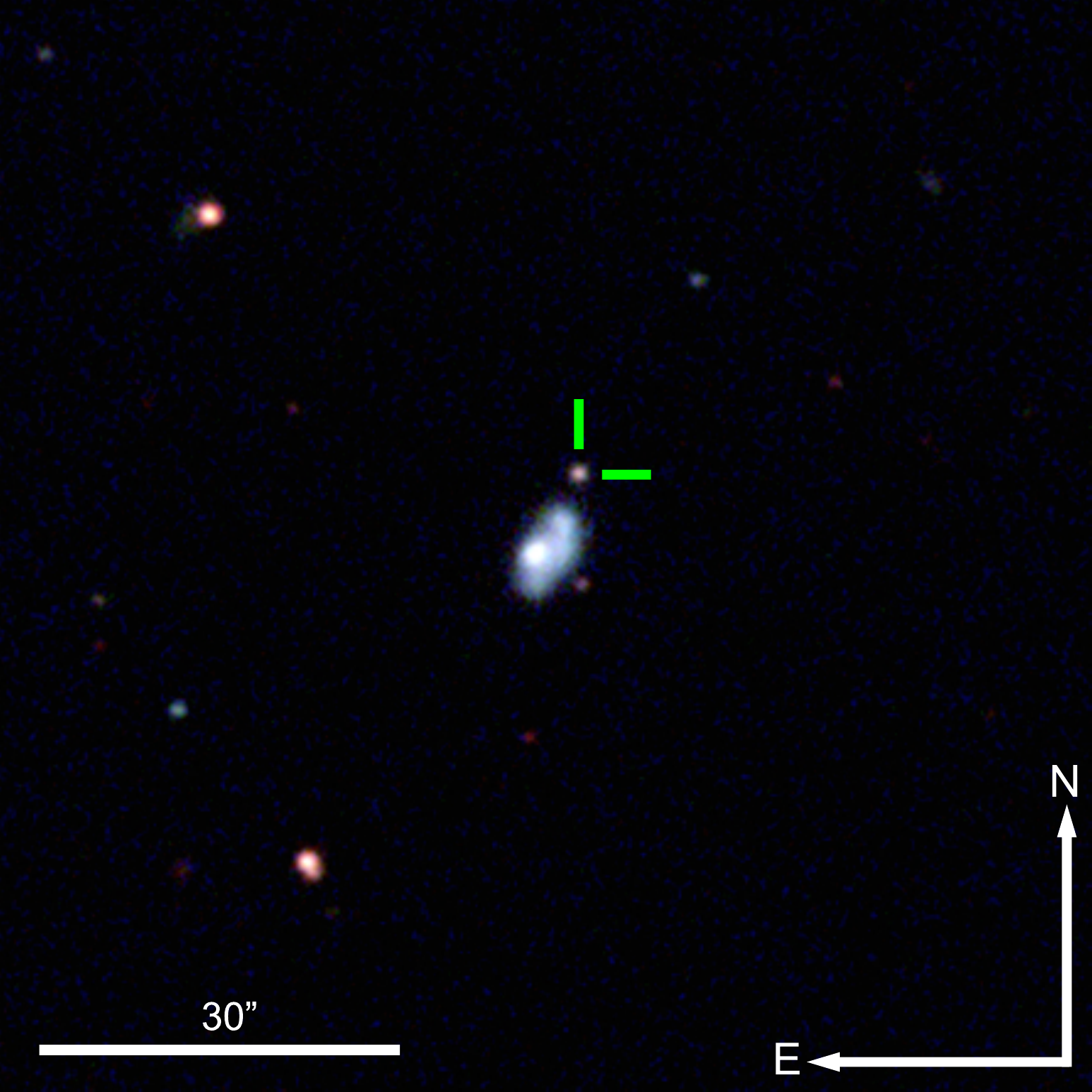}
	\caption{{$ 1.5'\times1.5' $ \textit{Bgri}-band composite image from the Liverpool Telescope observed 131 days after explosion showing \sn\ in the host galaxy \host.}}
	\label{fig:id}
\end{figure}

\sn/\hbox{SN~2018gk} (J2000 coordinates $\alpha = 16^{\rm h} 35^{\rm m} 54\fs60$, $\delta = +40\degr 01\arcmin 58\farcs01$) was discovered \citep{2018TNSTR..54....1B} in the galaxy \host\ (see Fig.~\ref{fig:id} for an image of the SN and host galaxy) by the All-Sky Automated Survey for Supernovae \citep[ASAS-SN;][]{2014ApJ...788...48S,2017PASP..129j4502K}. \sn\ was first detected by ASAS-SN on 2018-01-12.5 (UT dates are used throughout this paper) at a \textit{g}-band magnitude of 16.8. Based on the last nondetection on 2018-01-11.7 at a limit of $ V= 17.6 $\,mag, we choose an explosion epoch of 2018-01-12.1 ($\rm\EpEpoch\pm0.4$).

We adopt a total line-of-sight reddening of $ \ebv=0.0086\pm0.0011 $\,mag \citep{2011ApJ...737..103S}, which is entirely due to the Milky Way. We neglect any host-galaxy extinction owing to the absence of any \Nai~D absorption at the host redshift in all our spectra, indicating a very low or negligible contribution from the host galaxy. Assuming $ R_V=3.1 $, this corresponds to $ A_V=0.027\pm0.003 $\,mag.

The host \host\ is a late-type galaxy \citep{2015ApJ...800...80L} at a redshift of $ 0.031010 \pm 0.000005 $ \citep{2017ApJS..233...25A}. This gives a luminosity distance of $ D_L=140.5\pm2.3 $\,Mpc assuming a standard \textit{Planck} cosmology \citep{2016A&A...594A..13P}.

We used ultraviolet (UV) to mid-infrared data from {\it Galex}, the Sloan Digital Sky Survey (SDSS), and {\it WISE} to fit the spectral energy distribution (SED) with the {\tt FAST} code \citep{2009ApJ...700..221K} and estimate the properties of the host galaxy. We derived a host stellar mass of $\hbox{log}(M_*/\msun)=8.98^{+0.09}_{-0.08} $ and a specific star formation rate of $\rm log({\rm sSFR})=-9.83_{-0.03}^{+0.10} $.
The oxygen abundance in the MPA-JHU catalog \citep{2004MNRAS.351.1151B} is $\rm 12+log(O/H)=8.6 $ on the \cite{2004ApJ...613..898T} scale, implying roughly solar metallicity. The host properties  of \sn\ are typical of ccSN hosts \citep[see, e.g.,][]{2012ApJ...759..107K}. 
This SDSS oxygen abundance is based on a spectrum centred on the host's apparent nucleus, so the metallicity at the site of the SN is likely much lower since its projected separation of $\sim 5$\,kpc is outside the optical disk (see Fig.~\ref{fig:id}).

\begin{figure*}
	\centering
	\includegraphics[width=0.75\linewidth]{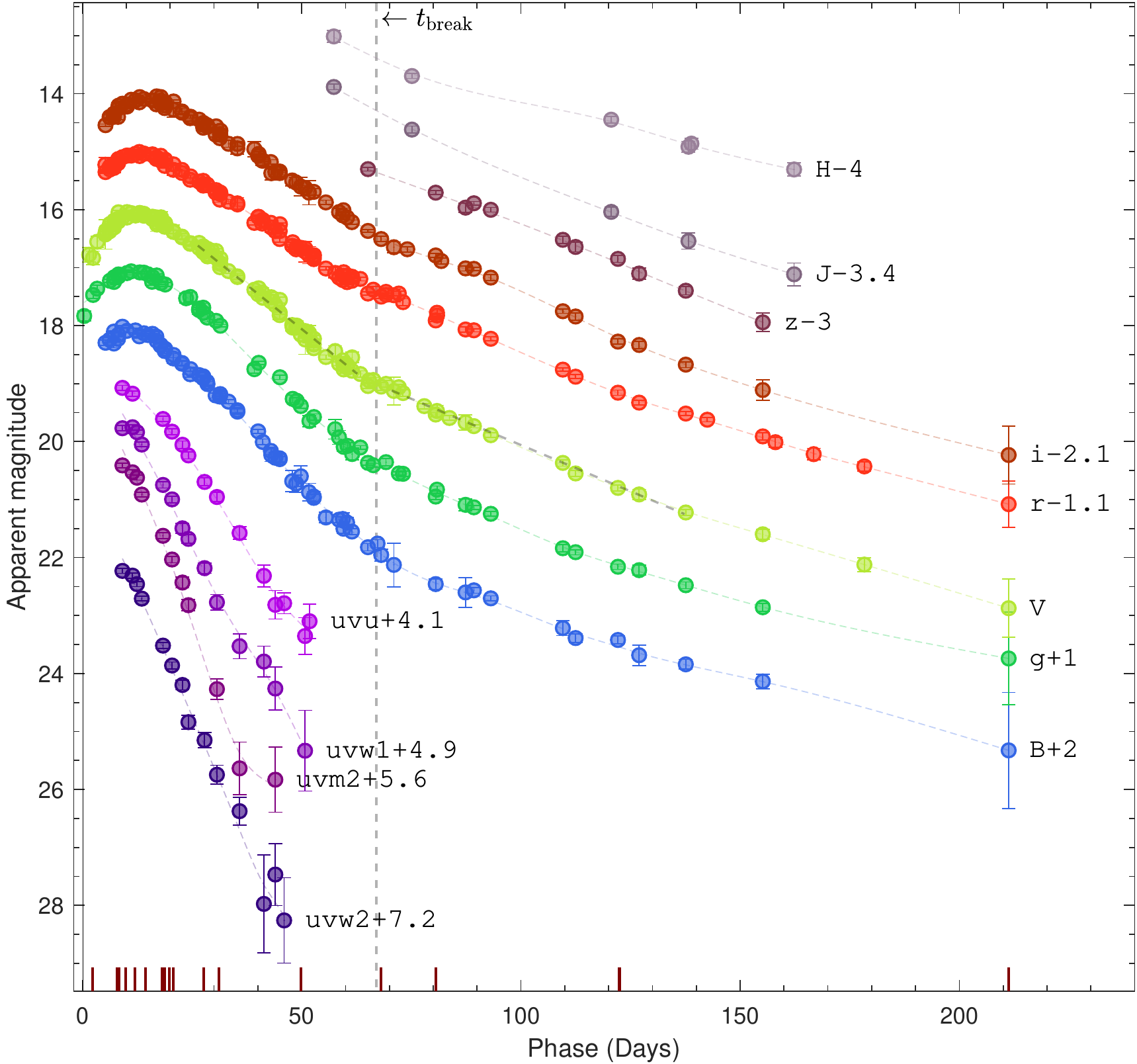}
	\caption{The photometric evolution of \sn\ in the UVOT-NUV, optical \textit{BVgriz}, and NIR bands. Epochs of spectral observations are marked by vertical bars at the bottom. The data points at day 211 are synthetic magnitudes computed from the spectrum taken at that epoch. {A pair of lines (grey colour) are overplotted on the $V$-band light curve to help illustrate the break in its slope.} The time of a break in the \lc\ slope is shown by a grey vertical dashed line, $ t_{\rm break}$.}
	\label{fig:all_lc}
\end{figure*}

\section{Observations and data} \label{sec:data}

Multiband photometric and spectroscopic observations were initiated soon after the discovery and were continued for 218 days (observer frame). Optical photometric data were obtained with the ASAS-SN quadruple 14\,cm ``Brutus" telescope, the Las Cumbres Observatory 1.0\,m telescope network \citep[LCOGT;][]{2013PASP..125.1031B}, the 0.6\,m telescopes at Post Observatory SRO (CA, USA) and Post Observatory Mayhill (NM, USA), the 0.5\,m DEdicated MONitor of EXotransits and Transients \citep[DEMONEXT;][]{2016SPIE.9906E..2LV,2018AAS...23131402V}, the 0.5\,m Iowa Robotic Telescope (both at the Winer Observatory, AZ, USA), and the 2.0\,m Liverpool Telescope (LT) at La Palma. Near-infrared (NIR) photometric observations were obtained with NOTCAM on the 2.6\,m Nordic Optical Telescope (NOT) at La Palma and WFCAM mounted on the 3.8\,m United Kingdom Infra-Red Telescope (UKIRT) at Maunakea. We also triggered near-ultraviolet (NUV) observations with the \textit{Neil Gehrels Swift Observatory} \citep{2004ApJ...611.1005G} Ultraviolet/Optical telescope (UVOT).

For optical and NIR data, point-spread-function (PSF) photometry was performed using a PyRAF-based pipeline\footnote{https://astro.subhashbose.com/software/diffphot} employing standard \daophot-\iraf\ photometry packages. The PSF radius and sky annulus were selected based on the mean full width at half-maximum intensity (FWHM) of stellar profiles for each image frame. Optical photometric calibration was done using SDSS \citep[for the \textit{g, r} and \textit{i} bands;][]{2017ApJS..233...25A} and APASS DR9 \citep[for the \textit{B} and \textit{V} bands;][]{2016yCat.2336....0H} catalogues of stars in the field. 2MASS \citep{2006AJ....131.1163S} local standards were used to calibrate the NIR photometry. No host-galaxy subtraction has been done for the optical or NIR data, as the SN is located outside the optical disk of the host and the contamination is negligible during our observed epochs. The UVOT photometry was performed with the UVOTSOURCE task in the HEAsoft package using an aperture of $ 5'' $ radius and placed in the Vega magnitude system, using the calibration from \cite{2011AIPC.1358..373B}. UVOT templates images observed at +426\,d were used to subtract the host-galaxy contamination in the lower resolution \textit{Swift} images. The photometric data for \sn\ are in Table~\ref{tab:photsn}. The uncertainties from the PSF photometry, differential photometry, and zero-point calibration are all propagated into the final reported uncertainties.

We obtained 17 optical spectra spanning from 2 to 218 days (in the observer's frame) after discovery. Long-slit spectroscopic observations were carried out using the FAST spectrograph \citep{1998PASP..110...79F} mounted on 1.5\,m Tillinghast telescope at the F. L. Whipple Observatory (AZ, USA), ALFOSC on the 2.6\,m NOT at La Palma, the Kast double spectrograph mounted on the 3\,m Shane telescope at Lick Observatory (CA, USA), LRIS \citep{1995PASP..107..375O} on the 10\,m Keck-I telescope at Maunakea (HI, USA), the Double Spectrograph (DBSP) on the 5\,m Hale Telescope at Palomar Observatory (CA, USA), BFOSC mounted on the Xinglong 2.16\,m telescope of the National Astronomical Observatories (CAS, China), OSMOS \citep{2011PASP..123..187M} on the 2.4\,m Hiltner Telescope at MDM Observatory (AZ, USA), OSIRIS on the 10.4\,m Gran Telescopio Canarias (GTC) at La Palma (Spain), and MODS \citep{2010SPIE.7735E..0AP} on the twin 8.4\,m LBT at Mount Graham International Observatory (AZ, USA). Spectra are obtained with the slit along the parallactic angle \citep{1982PASP...94..715F} in order to obtain accurate relative spectrophotometry. The Keck/LRIS spectrum was taken with an atmospheric dispersion compensator.

The medium-resolution spectra from MODS were reduced using the \textrm{modsIDL} pipeline, and the ALFOSC data using \texttt{ALFOSCGUI}\footnote{http://sngroup.oapd.inaf.it/foscgui.html; developed by E. Cappellaro}. The DBSP, OSMOS, and OSIRIS data were reduced with PyRAF-based \texttt{SimSpec}\footnote{https://astro.subhashbose.com/simspec/} pipeline. The standard FAST pipeline was used for the FAST spectra, with Massey standards \citep{1990ApJ...358..344M} for spectrophotometric calibration. The LRIS spectrum was reduced using the IDL-based \texttt{LPipe} pipeline \citep{2019PASP..131h4503P}, and the BFOSC spectra were reduced using standard \iraf\ routines. Kast data were reduced following standard techniques for CCD processing and spectrum extraction utilising \iraf\ routines and custom Python and IDL codes\footnote{https://github.com/ishivvers/TheKastShiv}. The spectroscopic observations are summarised in Table~\ref{tab:speclog}.

\sn\ was observed in photon-counting mode with the \swift\ X-ray Telescope \citep[XRT;][]{2005SSRv..120..165B} from 2018-01-21 (+9\,d) to 2018-03-06 (+52\,d) for a total exposure time of 35.2\,ks. The SN was also observed by the \textit{Chandra} X-ray Observatory on 2018-04-21 (+60\,d) with a 10\,ks exposure. Source counts were extracted in a circle with a radius of $24''$. 
Background counts were selected from a nearby source-free circular region. 
The Galactic column density in the direction of \sn\ is $N_H = 9.7\times 10^{19}$\,cm$^{-2}$ \citep{2005A&A...440..775K}.
The X-ray spectrum can be fit with a single absorbed power-law model with a photon index $\Gamma = 0.77^{+0.70}_{-0.75}$. Based on this model the count rates are converted into a $0.3-10$\,keV flux. The X-ray detections and upper limits are listed in Table.~\ref{tab:xray_det}.

There are two X-ray sources detected by the \textit{Swift} XRT in the vicinity of the optical position of \sn. The eastern one, which is also clearly detectable in the {\it Chandra} image, corresponds to a $ z=0.95 $ quasar (WISEA J163556.97+400138.6). The astrometric position (J2000) of the other X-ray source is $\alpha = 16^{\rm h} 35^{\rm m} 53\fs89$, $\delta = +40\degr 01\arcmin 58\farcs7$, with an uncertainty radius of 12\farcs3 (90\% confidence).
This position is 2\farcs5 away from the optical position of \sn\ and is well within the uncertainty radius. Other than the SN, no other object is visible in the UVOT images at that X-ray position, and given that the optical position is within the extraction radius of the X-ray source we are confident that these are X-rays emerging from the supernova and not from a background object (such as the eastern source). 

X-rays were only detected from +11\,d to +14\,d with luminosities in the range $\sim$ (4--6) $\times 10^{41}$\,\ergs. From +18\,d onward, the SN was no longer detected in the \swift\ observations. The \textit{Chandra} observation at +60\,d also could not detect any emission down to a limit of $ 3.8 \times 10^{40}\,\ergs $.

\section{Light curve} \label{sec:lightcurve}
Figure~\ref{fig:all_lc} shows the NUV, optical, and NIR light curves of \sn. We estimate a rise-to-peak time of $12.3\pm4.6$\,d for the \textit{V}-band light curve by fitting a third-order polynomial. This rise time is similar to that of some fast declining SNe~IIP/L (e.g., SN~2013ej, \citealt{2014MNRAS.438L.101V}; SN~2014G, \citealt{2016MNRAS.455.2712B}). After the peak, the optical light curves decline monotonically, and can be basically described by two-piece linear (in mag) components with a break near 65\,d (shown by a vertical dashed line, $ t_{\rm break} $). The early-time (20--65\,d) \lc\ decline rates are $ 17.1$ (\textit{uvw2}), $ 21.0 $ (\textit{uvm2}), $ 14.1 $ (\textit{uvw1}), $ 11.4 $ (\textit{uvu}), $ 7.5 $ (\textit{B}), $ 6.9 $ (\textit{g}), $ 6.0 $ (\textit{V}), $ 4.9 $ (\textit{r}), and $ 4.9 $  (\textit{i}) \maghundred\ for each band.
The decline rates during this photospheric phase are significantly steeper than those of normal SNe~II, including the fast-declining SNe~IIL \citep[see, e.g.,][]{2016MNRAS.455.2712B}. After the break near day 65, the light curve settles onto a slower declining tail until the end of our photometric observations at 178\,d. The slopes of this tail phase are $ 2.4 $ (\textit{B}), $ 2.7 $ (\textit{g}), $ 2.9 $ (\textit{V}), $ 2.8 $ (\textit{r}), $ 3.2 $ (\textit{i}), $ 3.0 $ (\textit{z}), $ 2.9 $ (\textit{J}), and $ 1.9 $ (\textit{H}) \maghundred\ for each band. These decay rates are significantly steeper than for a fully $ \gamma $-ray-trapped \nickel$ \rightarrow $\cobalt$ \rightarrow $\iron\ powered light curve with a slope of $ 0.98 $\,\maghundred.

\begin{figure*}
	\centering
	\includegraphics[width=0.8\linewidth]{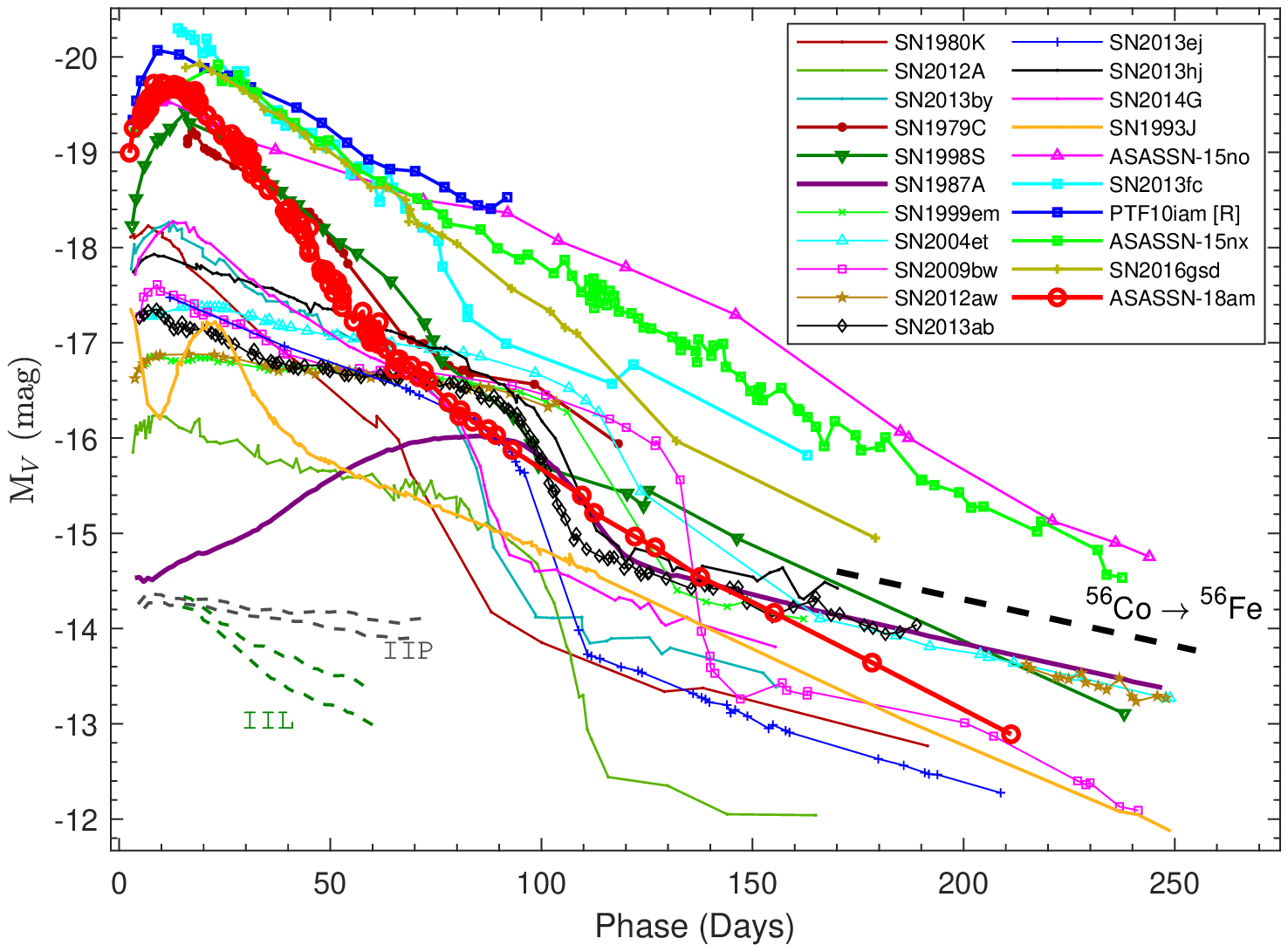}
	\caption{The absolute $ V $-band light curve of \sn\ compared with other normal and luminous H-rich SNe.
		The slope for a \cobalt$ \rightarrow $\iron\ radioactive decay law with full $ \gamma $-ray trapping is shown with a thick black dashed line. On the bottom-left side, pairs of grey and green dashed lines show the slope range for the SN~II-P and SN~II-L templates given by \citet{2014MNRAS.445..554F}. The adopted explosion time in $ \rm JD-2,400,000 $, 
		distance in Mpc, total \ebv\ in mag, and references for the light curves are as follows.
		SN 1979C  -- 43970.5, 16.0, 0.31, \citet{1982A&A...116...43B,1981PASP...93...36D}; 
		SN 1980K  -- 44540.5, 5.5, 0.30,  \citet{1982A&A...116...35B}; 
		SN 1987A  -- 46849.8, 0.05, 0.16,  \citet{1990AJ.....99.1146H};
		SN 1999em -- 51475.6, 11.7, 0.10, \citet{2002PASP..114...35L,2003MNRAS.338..939E};
		SN 2004et -- 53270.5, 5.4, 0.41, \citet{2006MNRAS.372.1315S}; 
		SN 2009bw -- 54916.5, 20.2, 0.31, \citet{2012MNRAS.422.1122I}; 
		SN 2012A  -- 55933.5, 9.8, 0.04, \citet{2013MNRAS.434.1636T};
		SN 2012aw -- 56002.6, 9.9, 0.07, \citet{2013MNRAS.433.1871B};
		SN 2013ab -- 56340.0, 24.0, 0.04, \citet{2015MNRAS.450.2373B};
		SN 2013by -- 56404.0, 14.8, 0.19, \citet{2015MNRAS.448.2608V};
		SN 2013ej -- 56497.3, 9.6, 0.06, \citet{2015ApJ...806..160B};
		SN 2013hj -- 56637.0, 28.2, 0.10, \citet{2016MNRAS.455.2712B};
		SN 2014G --  56669.7, 24.4, 0.25, \citet{2016MNRAS.455.2712B}; 
		ASASSN-15no -- 57235.5,153.5, 0.045, \citet{2018MNRAS.476..261B};
		SN 1993J  -- 9074.0, 3.68, 0.069,  \citet{1996AJ....112..732R}; 
		PTF10iam --  55342.7, 453.35, 0.19, \citet{2016ApJ...819...35A};
		SN 2013fc -- 56516.7,83.2, 0.935, \citet{2016MNRAS.456..323K};
		ASASSN-15nx -- 57219.1,127.5, 0.07, \citet{2018ApJ...862..107B}; and
		SN 2016gsd --  57648.5,311.6, 0.08, \citet{2020MNRAS.493.1761R}.}
	\label{fig:mv_comp}
\end{figure*}

\subsection{Absolute magnitude and bolometric luminosity and NIR colour} \label{sec:absM_bol}
The peak \textit{V}-band absolute magnitude of \sn\ is $ M_{V, {\rm peak}} = -19.70\pm0.27 $\,mag,
which lies between that of typical ccSNe and SLSNe, making \sn\ one of the small number of LSNe-II discovered thus far. In Figure~\ref{fig:mv_comp} we compare the absolute \textit{V}-band light curve of \sn\ with a sample of SNe~II comprised of normal SNe~IIP/L, SNe~IIb, and LSNe-II having peak absolute magnitudes similar to that of \sn. The peak absolute magnitude of \sn\ is brighter by $ \sim 1.4 $\,mag than the brightest of the normal SNe~II (e.g., SNe~1980K, 2013by, and 2014G), all of which are also fast-declining SNe~IIL. The early-time light curve of \sn\ has a faster decline rate than any SN in the comparison sample. For example, among the normal SNe~II, SN 1980K has one of the fastest \textit{V}-band photospheric-phase decline rates of $ \sim3.9 $\,\maghundred, while \sn\ declines at $ 6.0 $\,\maghundred. 
 
\sn\ exhibits a break in its \lc\ slope at $ \sim65 $\,d and thereafter settles onto a relatively slowly declining tail phase. Unlike normal SNe~II, which always show a drop of few magnitudes during the transition at $ \sim100 $\,d from the photospheric to the radioactive-decay phase, \sn\ does not have any such feature. ASASSN-15nx is another LSN-II lacking this transition phase, but it differs from \sn\ by having a continuous linear decline without any break demarcating the photospheric and the nebular phases.
The tail \lc\ decline rate of \sn\ ($ 2.9 $\,\maghundred) is also significantly steeper than that expected from a light curve powered by fully $ \gamma $-ray-trapped radioactive decay (slope 0.98\,\maghundred). However, its decline rate is comparable to that of other LSNe-II, such as ASASSN-15nx and also possibly SN~2016gsd.

\begin{figure}
	\centering
	\includegraphics[width=0.95\linewidth]{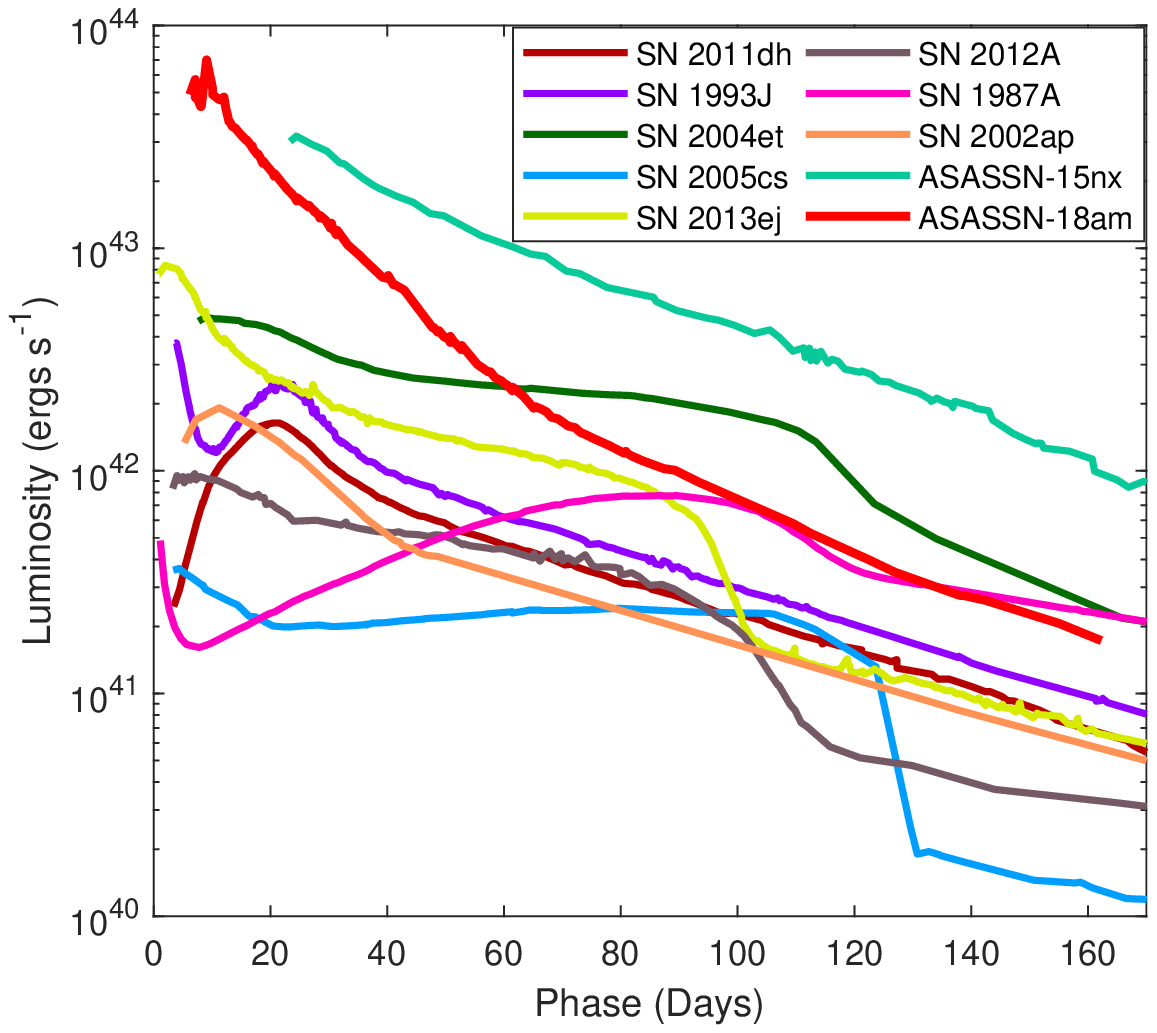}
	\caption{{The bolometric light curve of \sn\ compared with those of ASASSN-15nx \citep[LSN II;][]{2018ApJ...862..107B}, SN~2011dh \citep[IIb;][]{2015A&A...580A.142E}, SN~1993J \citep[IIb;][]{1996AJ....112..732R}, SN~2004et \citep[II;][]{2006MNRAS.372.1315S}, SN~2005cs \citep[II;][]{2006MNRAS.370.1752P,2009MNRAS.394.2266P}, SN~2013ej \citep[II;][]{2016MNRAS.461.2003Y}, SN~2012A \citep[II;][]{2013MNRAS.434.1636T}, SN~1987A \citep[II;][]{1990AJ.....99.1146H}, and SN~2002ap \citep[Ic-BL;][]{2003ApJ...592..467Y}. Comparison light curves for the normal-luminosity SNe are from \citet{2020arXiv200407244S}.}}
	\label{fig:bol}
\end{figure}

In Figure~\ref{fig:bol} we compare the bolometric light curve of \sn\ with that of other well-studied normal SNe and the LSN-II ASASSN-15nx. For \sn, the bolometric luminosities are calculated by fitting a blackbody to the NUV and optical bands until 54\,d. At later times, when NIR data ($J$ and $H$ bands) are available, we directly integrated over the SED, although blackbody models give very similar results (differences $ \lesssim0.05 $\,dex). At 57\,d, where we also have $K$-band data, we tested the blackbody model on the optical-NIR SED and found that the NIR fluxes are consistent with blackbody emission. For the direct integration of the SED, we assumed zero flux beyond $2000$\,\AA\ and extrapolated the $H$-band flux assuming a Rayleigh-Jeans approximation.
The NUV contribution to the bolometric luminosity becomes negligible after $\sim25$\,d. The bolometric light curve comparison shows trends similar to those for the absolute magnitudes, where the two LSNe-II \sn\ and ASASSN-15nx are roughly 1.5\,dex brighter  near maximum than typical ccSNe.

\begin{figure}
\centering
\includegraphics[width=\linewidth]{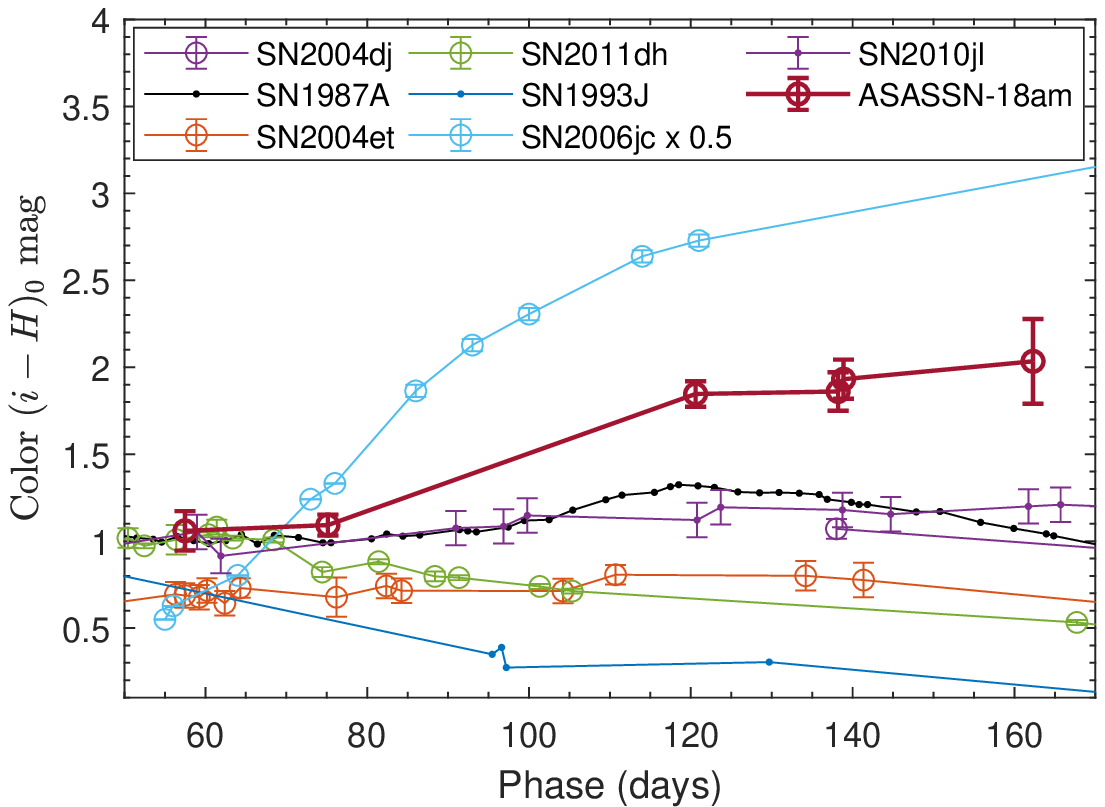}
\caption{{The extinction-corrected $(i-H)_0$ optical-to-NIR colour of \sn\ is compared with that of other SNe~II. SN~2006jc (Ibn) and SN~2010jl (IIn) are also included for comparison. For clarity, the colour curve of SN~2006jc is scaled down to half of its values.}}
\label{fig:nir_color}
\end{figure}

The $H$-band decline rate (see Fig.~\ref{fig:all_lc}) is significantly slower than all other bands (from $B$ through $J$), suggesting an excess in the $H$-band flux. In Figure~\ref{fig:nir_color} we compare the extinction-corrected $(i-H)_0$ NIR colour with a sample of SNe~II and SNe~IIb. SN~2006jc \citep[Ibn;][]{2008MNRAS.389..141M,2008MNRAS.389..113P} and SN~2010jl \citep[IIn;][]{2014ApJ...797..118F} are also included for comparison; they have strong IR excesses as a manifestation of dust emission.
Until $ \sim80 $\,d, \sn\ has a NIR colour similar to that of other SNe, and then it becomes significantly redder at later times. Owing to the lack of data redder than the $H$ band, the exact reason for the reddening is unclear. One possibility is the presence of some strong emission line in the $H$ band, or alternatively IR emission from warm dust. SNe can show IR emission from dust at longer wavelengths during late times (e.g., SN~1987A, \citealt[and references therein;]{1993A&A...273..451B}; SN~2004dj, \citealt{2011ApJ...732..109M}; SN~2004et \citealt{2010MNRAS.404..981M,2009ApJ...704..306K}; SN~2011dh \citealt{2015A&A...580A.142E}; and ASASSN-16at \citealt{2019ApJ...873L...3B}). SNe~2006j and 2010jl also showed strong excesses of flux in $H$ which were explained by IR emissions from dust \citep{2008MNRAS.389..141M,2014ApJ...797..118F}.

\section{Spectra} \label{sec:spectra}
\begin{figure*}
	\centering
	\includegraphics[width=0.8\linewidth]{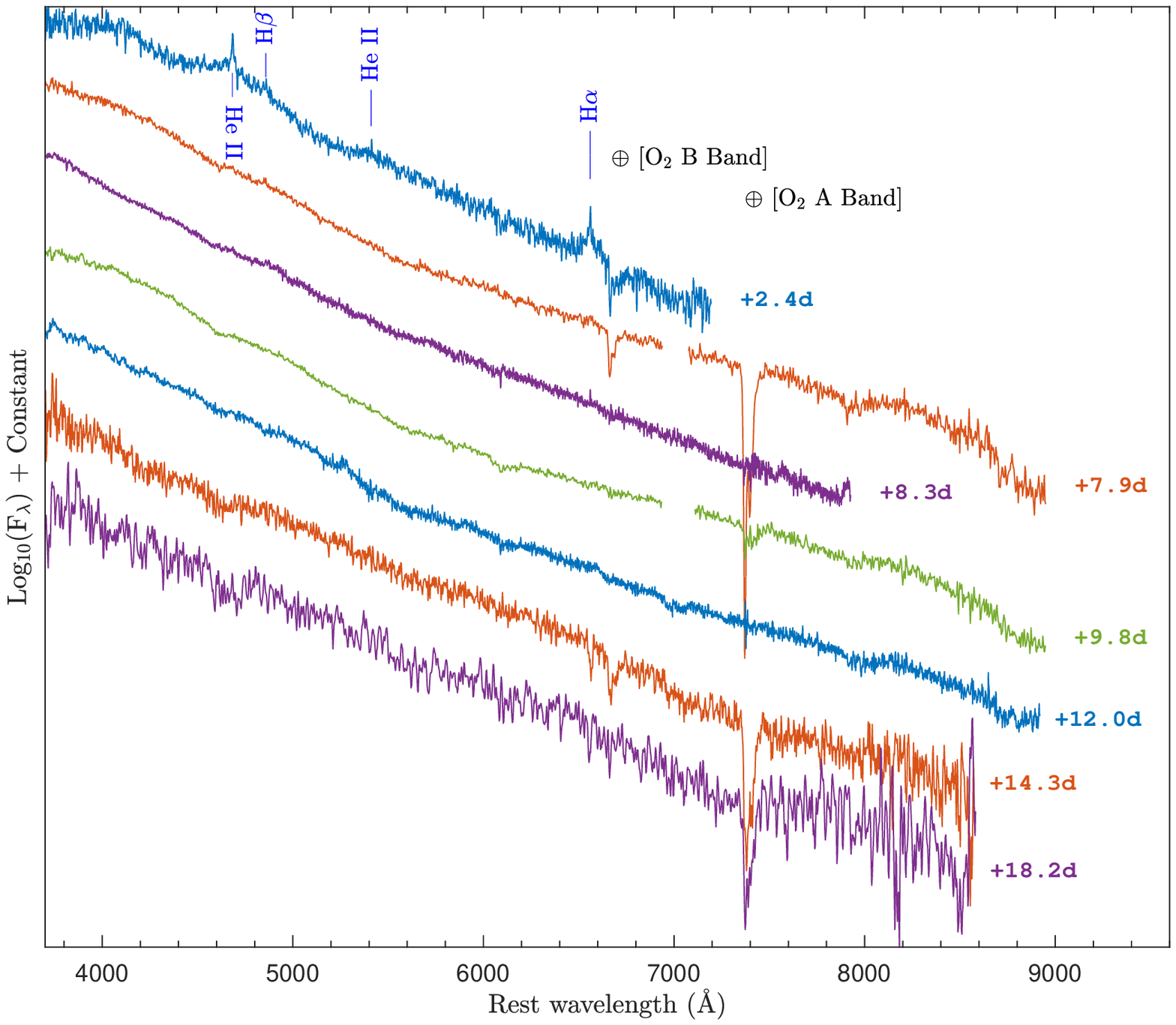}
	\caption{Spectral evolution of \sn\ from 2\,d to 18\,d showing flash-ionisation features in the first spectrum and a blue featureless continuum in the remainder. The positions for telluric absorption features are marked with $ \oplus $ symbol; they were not removed from some spectra.}
	\label{fig:all_spec_1}
\end{figure*}

\begin{figure*}
	\centering
	\includegraphics[width=0.8\linewidth]{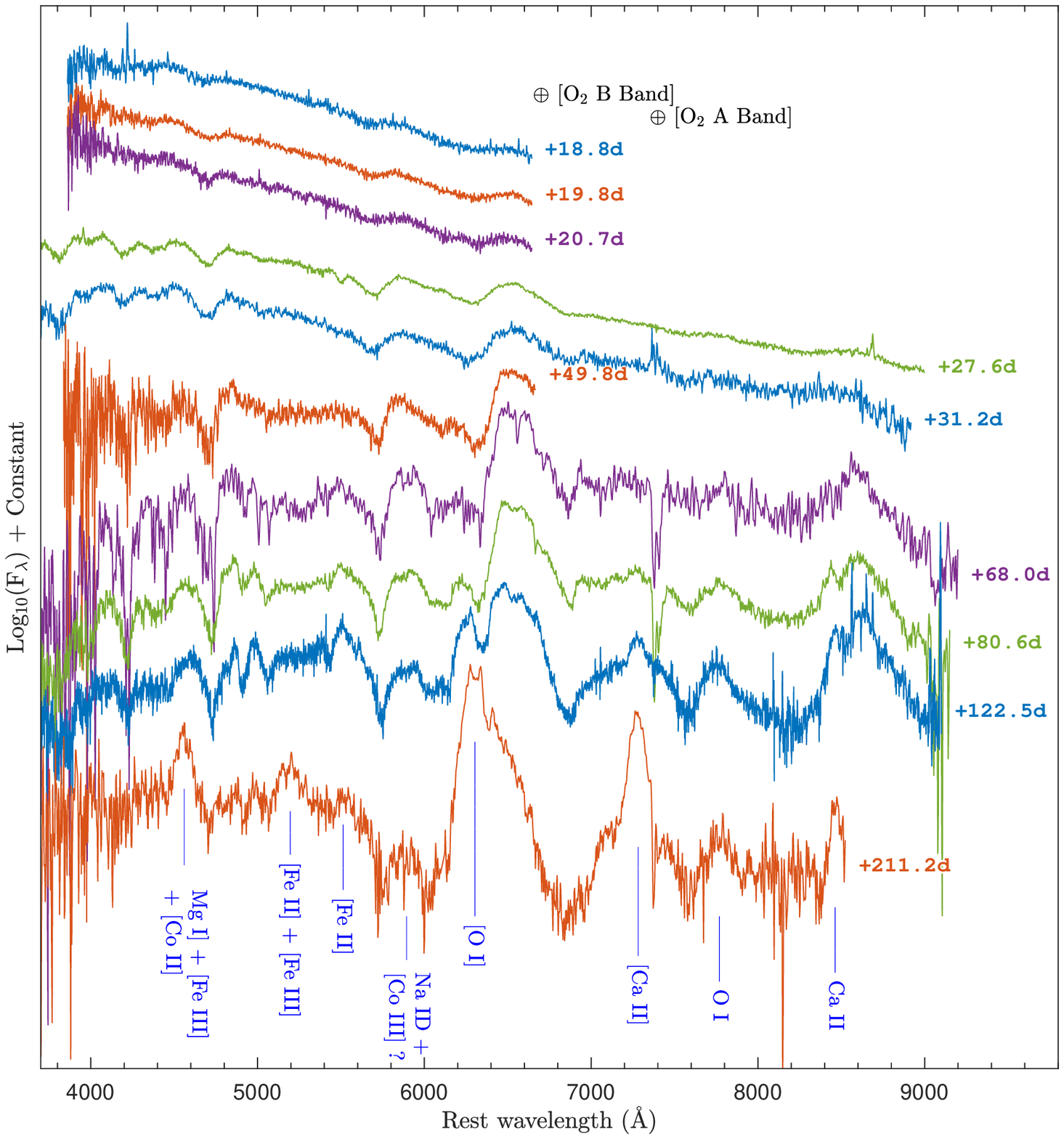}
	\caption{Spectral evolution of \sn\ from 19\,d to 211\,d showing the emergence and evolution of the spectral lines. The positions for telluric absorption features are marked with $ \oplus $ symbol.}
	\label{fig:all_spec_2}
\end{figure*}

Figures~\ref{fig:all_spec_1} and \ref{fig:all_spec_2} show the spectral evolution of \sn. Figure~\ref{fig:all_spec_1} displays the first 18\,d where the spectra are predominantly a featureless blue continuum. ``Flash-ionisation'' features are seen only in the earliest spectrum. The spectroscopic evolution from day 18 until the nebular phase is shown in Figure~\ref{fig:all_spec_2}.

\subsection{Flash ionisation and blue continuum} \label{sec:fi_bf}
The earliest +2.4\,d spectrum reveals narrow emission lines of H and ionised He on top of a blue continuum. These features originate from the recombination of CSM that was flash ionised by the initial shock-breakout radiation pulse \citep[see, e.g.,][]{1985ApJ...289...52N, 2014Natur.509..471G}. The emission-line profiles of \ha\ and \ion{He}{ii}~\ld4686 can be described by a combination of a broad Lorentzian and narrow Gaussian components, but the \hb\ and \ion{He}{ii}~\ld5411 profiles can be described by a single broad component; the narrow component cannot be detected in our low-resolution spectra. 
Such extended wings can be created by the radiative acceleration of the CSM by the shock-breakout luminosity \citep[e.g.,][]{2019MNRAS.483.3762K}. {Additionally, electron scattering can contribute to the formation of these extended wings \citep[e.g.,][]{2014ApJ...797..118F}.}
The measured FWHMs of the \ha\ profile are $ 2879\pm553 $\,\kms\ and $ 206\pm61 $\,\kms\ for the broad and narrow components (respectively), and $ 12,784\pm1236 $\,\kms\ and $ 876\pm110 $\,\kms\ for the broad and narrow components (respectively) of \ion{He}{ii}~\ld4686. However, the FWHM of the narrow component of \ha\ is limited by the instrumental resolution of the FAST spectrograph and should be considered an upper limit. For \hb\ and \ion{He}{ii}~\ld5411, the respective measured FWHMs for the broad components are $ 3319\pm1266 $\,\kms\ and $ 7971\pm1660 $\,\kms. 

The next spectrum, obtained at +7.9\,d, lacks any flash-ionisation features, leaving a blue and featureless continuum that lasts until +18.2\,d. We continued spectroscopy at a relatively higher cadence to constrain when emission lines start to appear. It was only 0.6\,d between the last featureless spectrum (+18.2\,d) and the appearance of spectral lines (+18.8\,d), though the +18.2\,d spectrum has significantly lower signal-to-noise ratio than the +18.8\,d spectrum. The presence of a blue continuum with or without the flash-ionisation features in the first 2--18\,d indicates a hot, optically thick envelope with temperatures of $ \sim11,000 $\,K to $ \sim15,000 $\,K. From a study of a sample of early-time SN~II spectra, \cite{2016ApJ...818....3K} suggested that flash-ionisation lines and featureless blue continua are more common in higher luminosity SNe, although none of the SNe in their sample is as luminous as \sn. \cite{2016ApJ...818....3K} also found that all flash-ionisation features in their sample are in spectra of age $ <10 $\,d. \sn\ is the only LSN-II having such early-time spectra. The earliest spectra of the well-studied LSNe-II PTF10iam \citep{2016ApJ...819...35A} and SN~2013fc \citep{2016MNRAS.456..323K} were taken at $\sim 15$\,d and show only a blue featureless continuum.
For other LSNe-II, such as ASASSN-15nx \citep{2018ApJ...862..107B} and SN~2016gsd \citep{2020MNRAS.493.1761R}, spectra were obtained only after $ >23 $\,d and already exhibit broad emission lines.

Under the assumption that the H$\alpha$ emission in the first 
spectrum is due to recombination of CSM photoionised by the 
shock-breakout radiation, we can estimate the wind mass-loss
rate. For this order-of-magnitude estimate, we will ignore
the extra complications caused by light-travel time effects \citep[see][]{2019MNRAS.483.3762K}. The H$\alpha$ recombination luminosity of
a fully ionised hydrogen wind is
\begin{equation}
L_{{\rm H}\alpha} = { \dot{M}^2 \alpha_{{\rm H}\alpha}\epsilon_{{\rm H}\alpha}
	\over 4 \pi v_w^2 m_p^2 R_{\rm in} },
\end{equation}
where $\dot{M}$ is the mass-loss rate, 
$\alpha_{{\rm H}\alpha} \approx 1.2 \times 10^{-13}$\,cm$^3$\,s$^{-1}$
is the Case B H$\alpha$ recombination rate, $m_p$ is the proton
mass, and $\epsilon_{{\rm H}\alpha}=1.8$\,eV. The inner edge of the wind is 
$R_{\rm in} \approx R_* + v_s t \approx 2 \times 10^{14}$\,cm assuming
a stellar radius of $R_* = 500\,{\rm R}_\odot$ and a shock speed of
$v_s = 10^4$\,\kms. {The mass-loss rate depends little on the assumed radius $R_*$ once $v_s t >> R_* $.} We assume a typical wind velocity of $ v_w=30$\,\kms\ since the observed line FWHM $ \approx 200 $\,\kms\ is limited by the instrumental resolution.
Given the observed luminosity of 
$L_{{\rm H}\alpha} = 9.8 \times 10^{38}$\,\ergs, we can solve
for the required mass-loss rate as
\begin{multline}
\dot{M} \approx 1.4 
\left[ {L_{{\rm H}\alpha} \over 10^{39}\,{{\rm erg}\,{\rm s}^{-1}} } \cdot 
{ R_{\rm in} \over 10^{14}\,\hbox{cm} } \right]^{1/2}
	\left[ v_w \over 30\,{{\rm km}\,{\rm s}^{-1}} \right]\\
	\times 10^{-4}\,{\rm M}_\odot\,\hbox{yr}^{-1} ,
\end{multline}
or $\dot{M} \approx 2 \times 10^{-4}\,{\rm M}_\odot\,\hbox{yr}^{-1}$ for our nominal values.

In the presence of density inhomogeneities, the actual mean wind density will be lower than estimated from the recombination luminosity. 
By the time of the second spectrum on day 7.9, the
inner radius would have expanded to $R_{\rm in} \approx 7 \times 10^{14}$\,cm
and we would expect the CSM emission to have dropped by a factor
of three. Combined with the increased continuum flux, it
makes the nondetection of flash-ionisation
features in this second spectrum plausible.

\subsection{Evolution of key spectral features}

Figure~\ref{fig:all_spec_2} shows the appearance and evolution of the spectral lines in \sn. At day 18.8, only 0.6\,d after the last featureless spectrum, broad P-Cygni profiles of \ha, \Hei~\ld5876, and \hb\ begin to appear and steadily strengthen. Lines of intermediate-mass and iron-group elements also appear after the $ +31 $\,d spectrum and persist until the last epoch of observation at $ +211 $\,d.  

Forbidden emission lines of [\Oi]~\ldld6300, 6364 and [\Caii]~\ldld7291, 7324, which are characteristic nebular-phase features, become prominent from $ +80.6 $\,d onward. During the nebular phase ($ >80 $\,d), the most dramatic evolution is seen at $ \sim 6000$--7000\,\AA. The apparent \ha\ emission becomes relatively weak compared to the metal lines, while the [\Oi] emission grows substantially stronger at $ +122.5 $\,d and is the dominant emission feature in the $ +211.2 $\,d spectrum. During the nebular phase, the \Hei\ emission component becomes stronger but it is blended with the weak \ha\ emission, as identified in the \synow\ models at earlier phases (see below). The weak and unresolved \ha\ emission in the nebular phase indicate a low hydrogen content in the ejecta. In the late-time spectrum at +211.2\,d, the broad feature is predominantly [\Oi] emission, while the extended blue wing is likely a blend of \Hei\ and [\Nii]~\ldld6548, 6583. This spectral feature is reminiscent of the nebular-phase spectra of SNe~IIb or SNe~Ib.

\begin{figure*}
	\centering
	\includegraphics[width=\linewidth]{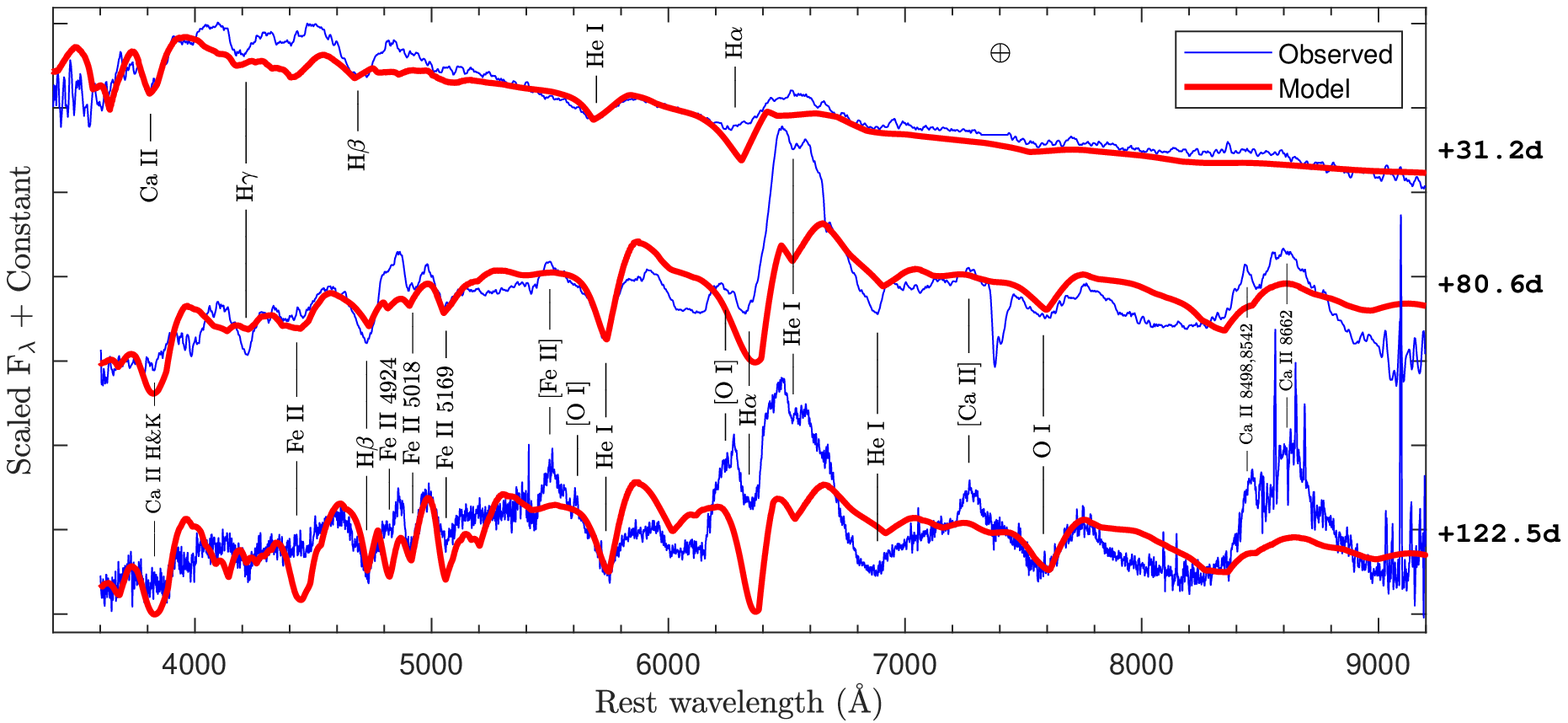}
	\caption{The \synow\ models and line identifications for \sn\ at three epochs. The observed spectra are corrected for extinction and redshift. All prominent permitted lines are labeled by marking the corresponding absorption component except for \Caii~\ldld8498, 8542, 8662, while the emission peaks of forbidden lines are marked. Forbidden lines were identified by referring to previous identifications in the literature. Note that telluric absorption was not removed from the +80.6\,d spectrum.}
	\label{fig:synow}
\end{figure*}

We used \synow\ \citep{2002ApJ...566.1005B,1997ApJ...481L..89F} to model the spectra and identify lines using a set of atomic species \Hi, \Hei, \Oi, \Feii, \Tiii, \Scii, \Caii, and \Baii. The models with the line identifications at three different phases are shown in Figure~\ref{fig:synow}. Although \synow\ is only suitable for modeling spectra during the photospheric phase, we also modeled the +122.5\,d spectrum, as it has only partly transitioned to the nebular phase; P-Cygni profiles are still visible with a photospheric velocity of $ \sim6500\,\kms $.

In comparison to normal SNe~II, \sn\ has more complex blends of lines, especially on the blue side of the spectrum ($ <5500 $\,\AA), which the \synow\ models cannot fully reproduce. However, we could identify the dominant species, among which the \Hei\ lines are one the most important identifications.  
In spectra older than a few weeks, the strong absorption profiles near $ 5700 $\,\AA\ are generally attributed to \Nai~D~\ldld5890, 5896 in H-rich SNe~IIP/L. However, this line is also very close to \Hei~\ld5876 which is difficult to distinguish from \Nai~D when the line velocities are high. Nevertheless, in the $+80.6$\,d and $+122.5$\,d spectra of \sn, we identify this line as \Hei\ instead of \Nai~D. Identified as \Hei, the line velocity is well aligned with all the other metallic line velocities as well as the photospheric velocity of the model, whereas fitting the feature as \Nai~D would require a $ \sim 30\% $ ($ \sim 2000 $\,\kms) higher velocity than the photosphere. Moreover, by invoking \Hei\ as the identification, we could also reproduce two additional absorption features --- one at $ \sim6900 $\,\AA\ (\Hei~\ld7065) and the other as a minor dip at $ \sim6560 $\AA\ (\Hei~\ld6678) near the top of apparent the \ha\ emission -- which further corroborates our line identification.

\subsection{Spectroscopic comparison} \label{sec:spec_compare}
\begin{figure}
	\centering
	\includegraphics[width=\linewidth]{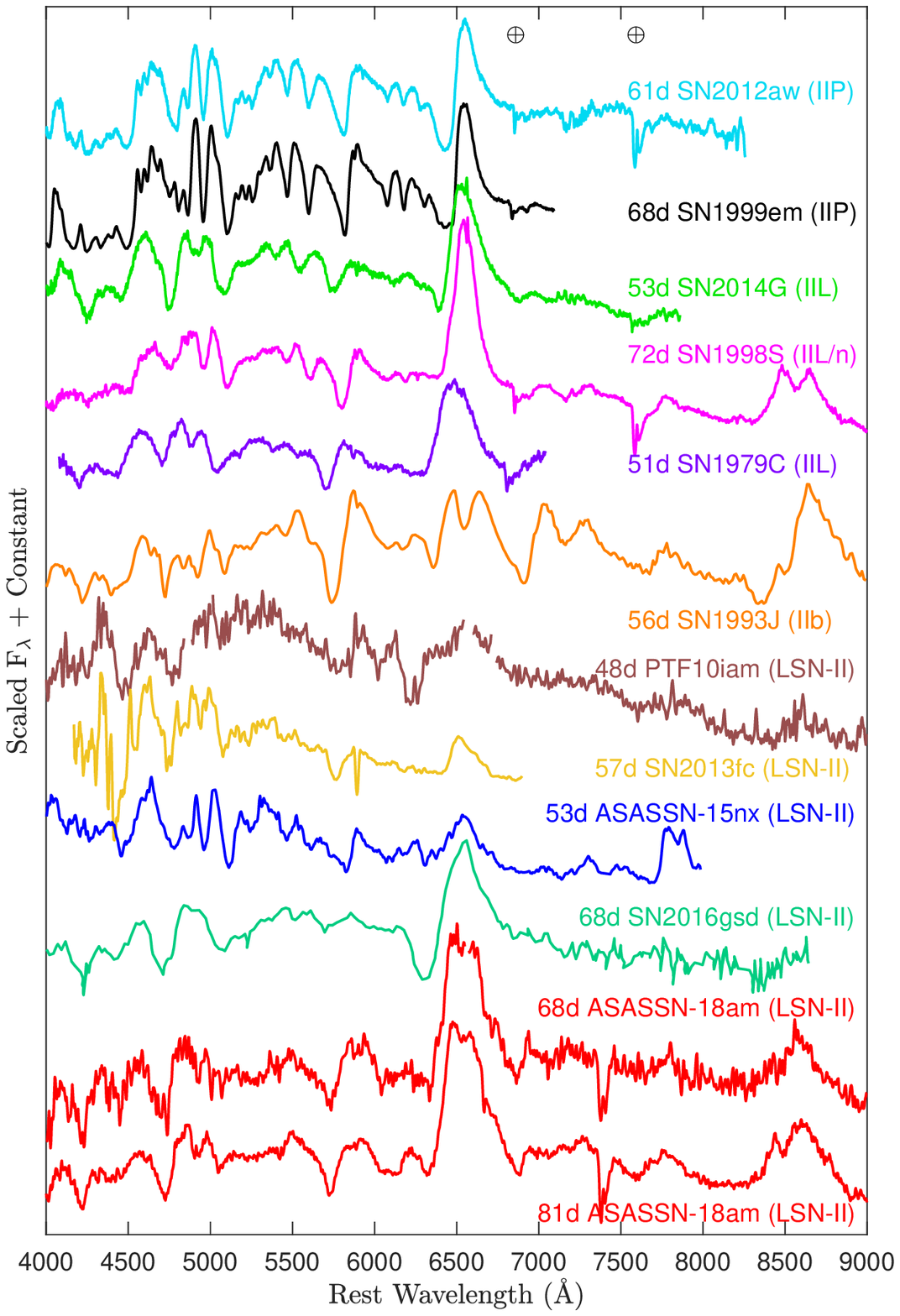}
	\caption{The 68~d and 81~d spectra of \sn\ compared with spectra of other SNe. The name, type, and phase of the SNe are labeled in the figure. Luminous Type II SNe like \sn\ are labeled as ``LSNe-II.'' The positions for telluric absorption features are marked with $ \oplus $ symbol.}
	\label{fig:sp_comp_d68}
\end{figure}

\begin{figure}
	\centering
	\includegraphics[width=\linewidth]{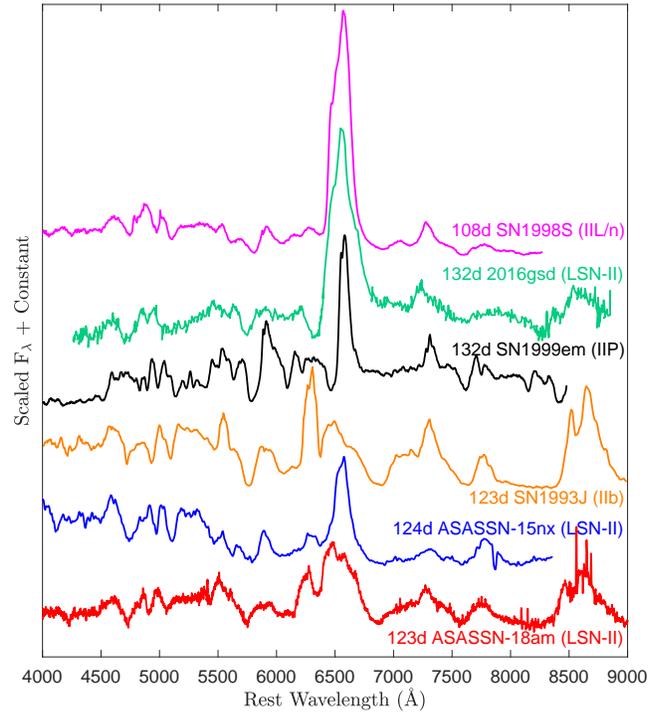}
	\caption{Same as Fig.\ref{fig:sp_comp_d68} but for the +123\,d spectrum of \sn.}
	\label{fig:sp_comp_d123}
\end{figure}

\begin{figure}
	\centering
	\includegraphics[width=\linewidth]{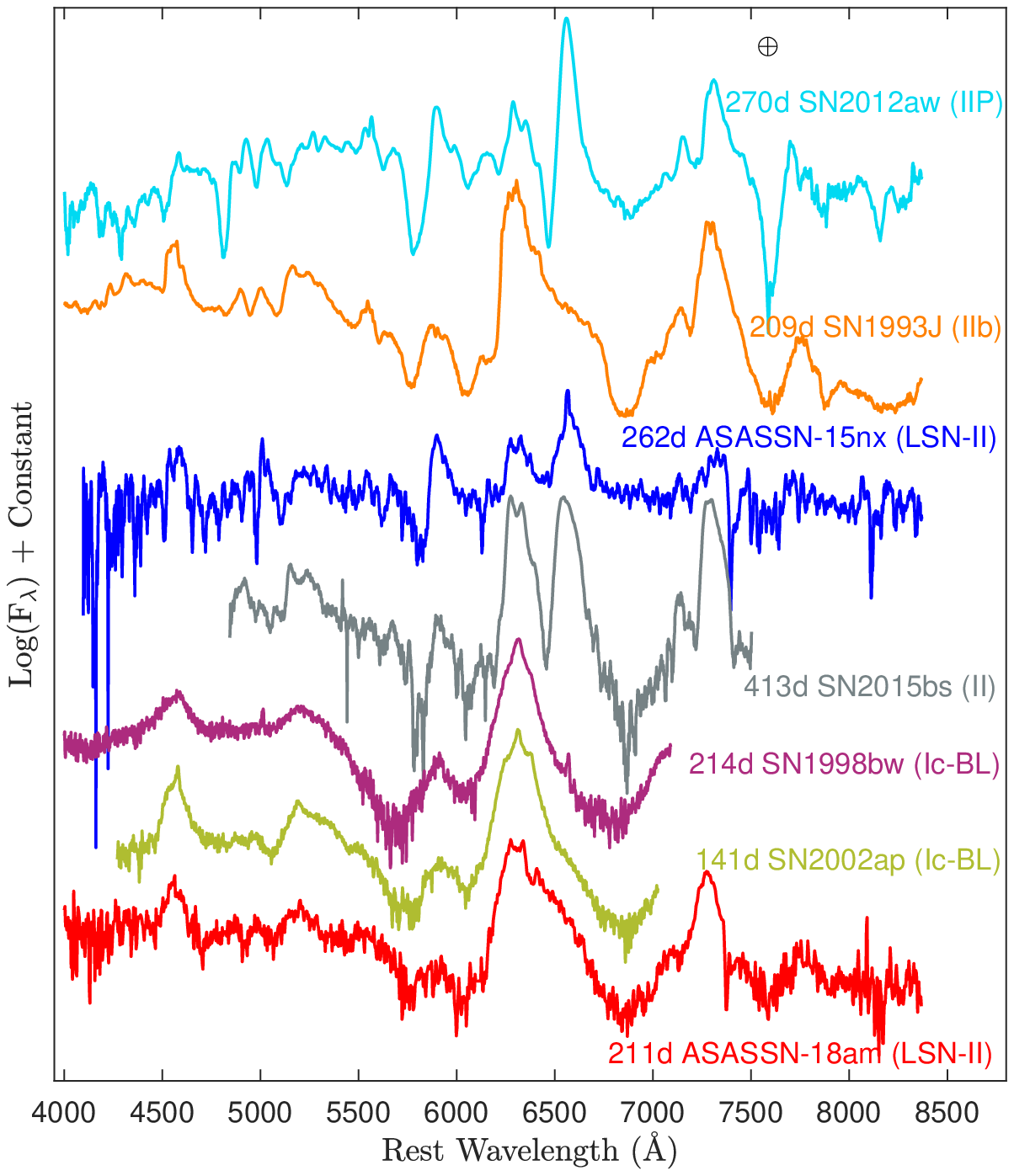}
	\caption{Same as Fig.\ref{fig:sp_comp_d68} but for the +211\,d spectrum of \sn\ and plotted on a logarithmic flux-density scale.}
	\label{fig:sp_comp_d211}
\end{figure}

In Figures~\ref{fig:sp_comp_d68} to \ref{fig:sp_comp_d211} we compare the spectra of \sn\ with those of other H-rich SNe, including the LSNe-II PTF10iam, SN~2013fc, ASASSN-15nx, and SN~2016gsd. In Figure~\ref{fig:sp_comp_d68} we compare the +68\,d and +81\,d spectra of \sn\ with other SNe at similar phases. Overall, \sn\ has many similarities to other SNe~IIP/L. An \ha\ profile with a weak absorption component is similar to that of other fast-declining SNe~II such as SN~1979C and SN~1998S. However, the apparently broad \ha\ profile near 6500\,\AA\ is identified as a blend of \ha\ and \Hei\ in \synow, and does not match the SN~IIP/L spectra. The partially blended \ha\ and \Hei\ profile of SN~IIb 1993J is somewhat similar to that of \sn, but the \ha\ and \Hei\ lines are not distinctly resolved in the latter. This suggests that even if \sn\ is spectroscopically a ``SN~IIb,'' it is likely richer in hydrogen than typical SNe~IIb.

In Figure~\ref{fig:sp_comp_d123} the early nebular spectrum of \sn\ at 123\,d is compared with a subset of the SNe from the previous figure. The \Feii\ and \hb\ lines near 4700\,\AA, the [\Caii] emission near 7300\,\AA, and the \Oi~\ld7774 line are similar to the comparison sample. However, the line profiles in the range 6000--7000\,\AA\ are significantly different from those of the other SNe, with SN~IIb 1993J being the closest match. This again seems to imply that the \ha\ in \sn\ is weaker than in SNe~IIP/L or ASASSN-15nx, but stronger than in SN~IIb 1993J. Strong [\Oi]~\ldld6300, 6364 emission is also seen in SNe~1993J and ASASSN-15nx, suggesting that these SNe have relatively thin hydrogen envelopes and enter the nebular phase much earlier than their H-rich counterparts (SNe~IIP/L). 

The nebular-phase spectrum of \sn\ at +211\,d is compared with other ccSNe in Figure~\ref{fig:sp_comp_d211}. The spectrum is again very similar to that of SN~IIb 1993J, especially the broad and blended feature near 6300\,\AA\ which is formed by [\Oi], \Hei, and [\Nii]. Unlike the previous spectrum, \ha\ is likely very weak or nonexistent at this phase and now [\Nii] has a stronger contribution \citep{2015A&A...573A..12J}. SN~2015bs, a SN~II from a massive ($ \sim25 $\,\msun) progenitor \citep{2018NatAs...2..574A}, is also included for comparison. It shows strong [\Oi] emission like \sn\ but with additional, prominent \ha\ emission. Typical SNe~II (e.g., SN~2012aw) and the LSN-II ASASSN-15nx show much weaker [\Oi] emission. 
We also included two SNe~Ic-BL, SN 1998bw \citep{2001ApJ...555..900P} and SN 2002ap \citep{2003PASP..115.1220F}, for comparison. The blue (4300--6000\,\AA) spectra of \sn\ are remarkably similar to these SNe~Ic-BL. Two notable similarities are the prominent blends of \ion{Mg}{i}]~\ld4750, [\ion{Fe}{iii}]~\ld4658, and [\ion{Co}{ii}]~\ld4624 near 4600\,\AA, and the blends of [\ion{Fe}{iii}]~\ld5270 and [\Feii] multiplets near 5200\,\AA\ (see \citealt{2007ApJ...670..592M} for the line identifications). Such strong lines of iron-group elements are not seen in other SNe~II. {The emission feature near 5900\,\AA\ resembles the \Nai~D~\ldld5890, 5896 doublet commonly seen in ccSNe nebular spectra. In earlier phases the same region is dominated by \Hei. However, owing to the presence of relatively strong iron-group lines, we also expect some contribution from [\ion{Co}{iii}]~\ld5888 emission in this feature. We measure the flux ratio between the 123\,d and 211\,d spectra to be $ \sim7.5\pm0.5 $, while for pure cobalt decay it is expected to be $ \sim10.4 $ \citep{2015MNRAS.454.3816C}. This re-affirms that the 5900\,\AA\ feature at these phases is dominated by \Nai~D or weak \Hei\ emission.}

\subsection{Line velocities} \label{sec:line_vel}

\begin{figure}
	\centering
	\includegraphics[width=\linewidth]{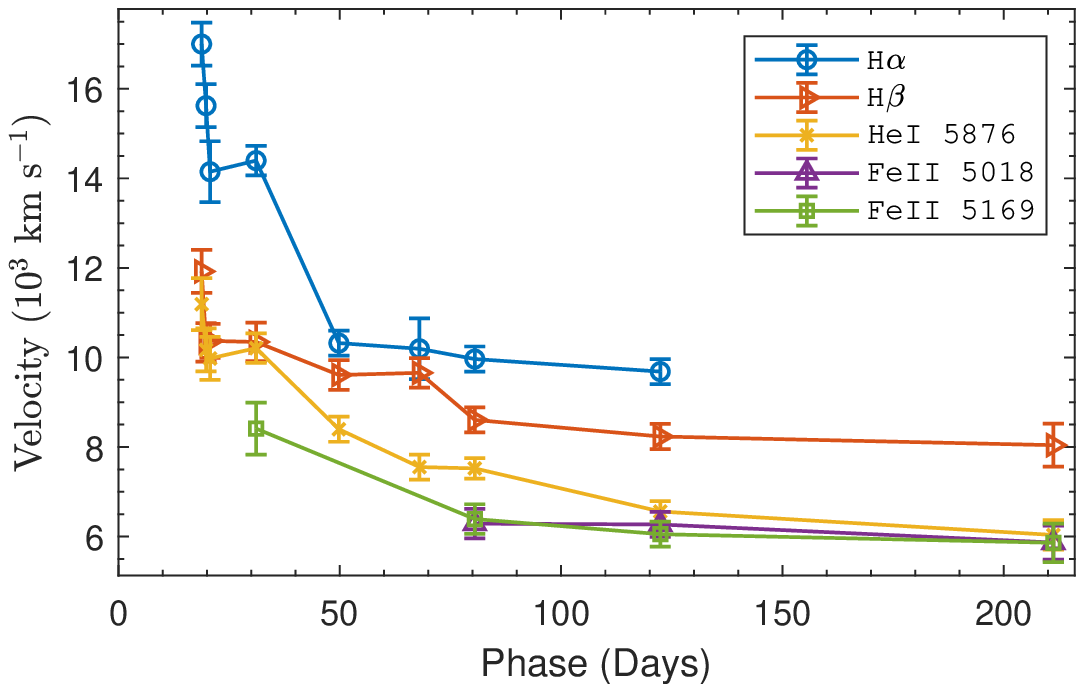}
	\caption{The velocity evolution of \ha, \hb, \Hei, and \Feii\ lines for \sn. }
	\label{fig:vel_profile}
\end{figure}

In Figure~\ref{fig:vel_profile} we show the \ha, \hb, \Hei, and \Feii\ line-velocity evolution defined by the minimum of the absorption feature. Broad P-Cygni profiles of \ha, \hb, and \Hei\ start to appear from +18.8\,d with high expansion velocities ($ \sim17,000 $\,\kms\ for \ha). Since this spectrum was only 0.6\,d after the last blue, featureless spectrum, the lines are likely formed very close to the outermost layer of the ejecta at high velocities. Over the first three epochs to +20.7\,d the velocities drop rapidly, and afterward they decline slowly. After +49.8\,d, the \ha\ and \hb\ velocities remain almost constant at $ \sim 10,000 $\,\kms\ and $ \sim 8500 $\,\kms, respectively. The flat velocity profiles indicate a stratified shell of \Hi\ with little or no mixing in the ejecta. The highest \Hei\ velocity of $ \sim11,000 $\,\kms\ is roughly at the lower bound of the \ha\ velocity. This suggests that, although \Hi\ is mostly confined to a shell, it is not detached from the \Hei\ core.

\begin{figure}
	\centering
	\includegraphics[width=\linewidth]{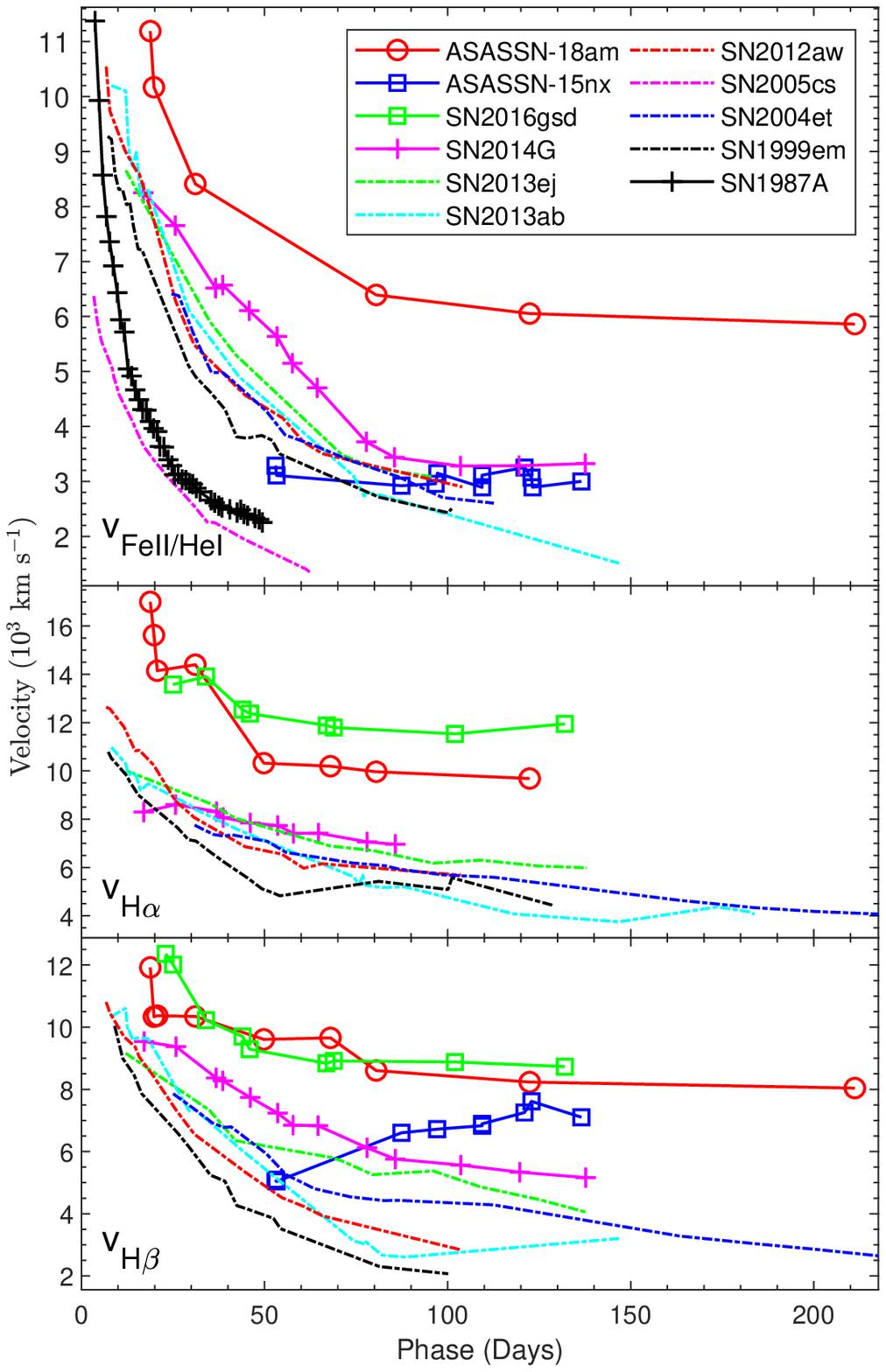}
	\caption{The top panel shows \Feii\ line velocity profiles as a proxy for photospheric velocity. \Hei\ velocities are subtitled in the profile when \Feii\ lines are not detectable during early phases. The references for the sources of spectra are the same as for the light curves in Figs.~\ref{fig:mv_comp} and \ref{fig:bol}, with the addition of \citet[SN~1987A]{1995ApJS...99..223P} and \citet[SN~2014G]{2016MNRAS.462..137T}.}
	\label{fig:vel_comparison}
\end{figure}

We compare the \ha, \hb, and \Feii/\Hei\ line velocities with a sample of other SNe~II in Figure~\ref{fig:vel_comparison}. The \Feii\ lines represent the photospheric velocity, and during early phases (+18.8\,d to +20.7\,d for \sn) when \Feii\ lines are not detectable the \Hei\ lines are a good proxy for photospheric velocity \citep{2006MNRAS.372.1735T,2014ApJ...782...98B}.
The comparison sample includes normal SNe~II with prominent plateaus (IIP; e.g., SNe 2004et, 1999em, 2012aw), fast-declining SNe~II (IIL; e.g., SN 2014G), intermediate decline rate SNe~II (e.g., SN 2013ej), and \sn-like LSNe-II (e.g., ASASSN-15nx, SN 2016gsd). 
SNe \sn\ and 2016gsd are among those with highest velocities, and their velocities are significantly higher than those seen in normal SNe~II.
The earliest \ha\ velocity of \sn\ at $ \sim17,000 $\,\kms\ is larger than for any other object both in our sample and in
the 122 SNe~II analysed by \cite{2017ApJ...850...89G} where the maximum \ha\ velocity is $ \sim15,000 $\,\kms. 
SNe with faster light-curve declines tend to show flatter and overall higher \Hi\ velocity evolution curves \citep{2014MNRAS.445..554F,2015ApJ...806..160B} as compared to SNe with more slowly declining light-curves. A similar trend is seen in Figure~\ref{fig:vel_comparison}, with the exception of ASASSN-15nx. \sn\ is the steepest declining SN~II, followed by SNe 2016gsd, 2014G, 2013ej, and then the rest of the SNe~IIP with slowly declining or nearly flat light curves.

\section{Light curve and its powering mechanism} \label{sec:lc_discussion}
The peak luminosity of \sn\ is about an order of magnitude higher than that of typical SNe~II. This makes it challenging to explain its powering mechanism. It is also one of the fastest declining SNe~II, with a decline rate of $ 6.0 $\,\maghundred. Here we discuss some theoretical models and their limitations. 
{We fit the models to the bolometric light curve computed in \S\ref{sec:absM_bol}.}

\subsection{Simple radiative diffusion} 
\label{sec:model_rad}
\begin{figure}
	\centering
	\includegraphics[width=\linewidth]{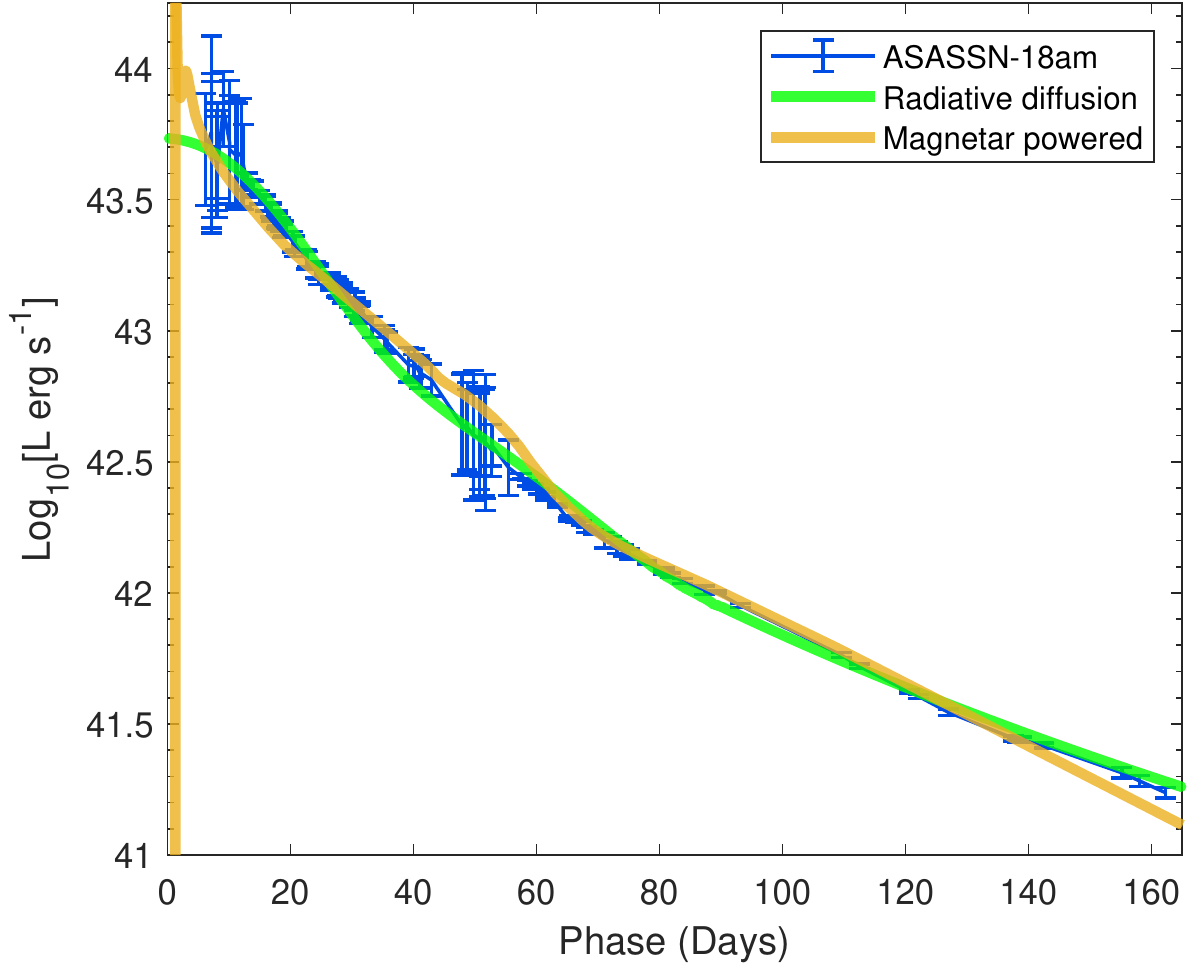}
	\caption{The \lc\ fits of \sn\ for the radiative diffusion and magnetar spin-down models.}
	\label{fig:models}
\end{figure}

Here we use semi-analytical models with adiabatically expanding ejecta combined with radioactive heating and undergoing diffusion cooling as originally outlined by \cite{1980ApJ...237..541A} and \cite{1989ApJ...340..396A}. We first tried a single-component implementation of the model as described by \cite{2015ApJ...806..160B}. 
{This model is unable to reproduce the steepness of the light curve earlier than $ \sim30 $\,d. Therefore, we extend this into a two-component ejecta model, where the core and a less massive envelope are treated independently, and the \nickel\ is confined to the core. This two-component formulation follows an approach similar to that of \cite{2016A&A...589A..53N}; however, the implementation of the radioactive energy-deposition function is slightly different. We accurately account for the positron energy deposition from \cobalt\ decay and also ensure that the loss due to $ \gamma$-ray leakage applies only to the \nickel\ heating and not to the radiated component of internal energy. These factors are important for steeply declining light curves having significant $ \gamma $-ray leakage.}
This best-fit model is shown with a solid green line in Figure~\ref{fig:models}. In this model, the ejecta become optically thin by $ \sim80 $\,d, and thereafter the tail is entirely powered by radioactive decay.

The best-fit model has a radioactive \nickel\ mass {of $ M_{\rm Ni}=0.43\pm0.06$\,\msun\ with a $\gamma$-ray trapping parameter of $ t_{0\gamma}=53\pm7 $\,d,} where $  t_{0\gamma} $ defines the time-dependent $ \gamma $-ray optical depth as $  \tau_\gamma \approx t_{0\gamma}^2/t^2 $ \citep{1999astro.ph..7015J}.
The radioactive decay power alone cannot account for the radiated energy. The model has an total ejecta mass of $ 3-4 $\,\msun\ with a very large kinetic energy $ E_{\rm kin} =$ (7--10) $\times 10^{51} $\,erg, and most of the energy is carried by the $ \sim1.5\,\msun $ envelope. This large kinetic energy is consistent with the high expansion velocities measured from spectra (see \S\ref{sec:line_vel}). 
The large derived \nickel\ mass and high kinetic energy are also consistent with the empirical correlation for ccSNe found by \citet[see their Fig.~3]{CITE3}. 

{We also estimated the \nickel\ parameters using the time-weighted luminosity integral method that conserves the energy considering adiabatic losses \citep{2013arXiv1301.6766K, ET, 2020arXiv200407244S} and found $ M_{\rm Ni}=0.43\pm0.05$\,\msun\ with $ t_{0\gamma}=51\pm4 $\,d. 
The corresponding ``ET'' value is $(8\pm1) \times 10^{55}$\,ergs, which is the excess energy released (time-weighted integral) in the ejecta without the radioactive energy deposition. The derived $ M_{\rm Ni}$ and $ t_{0\gamma}$ values are identical to those estimated from direct fitting of the radioactive energy deposition function used for the radiative diffusion model.}

\subsection{Magnetar spin-down} \label{sec:model_magnetar}
A second possible powering mechanism is the spin-down of a newly formed magnetised neutron star, thereby injecting additional energy into the ejecta. Such magnetar engines can produce SN light curves with a wide range of luminosities depending on the spin period and the strength of the magnetic field \citep[e.g.,][]{2010ApJ...717..245K}. Magnetars are often discussed as a plausible powering mechanism for superluminous SNe \citep[e.g.,][]{2017ApJ...850...55N,2018A&A...613A...5D}. A magnetar engine was also proposed as a possible powering source for the LSN-II ASASSN-15nx \citep{2019AstL...45..427C}.
A central magnetar engine can lead to bipolar geometry for the SN ejecta \citep[e.g.,][]{2017MNRAS.472..616S}, which can be seen as an asymmetry in the nebular emissions lines, like in [\Oi]. Such features are not seen for \sn, but this could simply be due to the viewing angle.

For the best-fit magnetar model (see Fig.\ref{fig:models}, dark-yellow line) the progenitor at the time of explosion has an envelope mass of $ \sim 2 $\,\msun\ and the kinetic energy of the explosion is $ E_{\rm kin}=3\times10^{51} $\,erg. The estimated mass of \nickel\ synthesised in the explosion is $ 0.28 $\,\msun\ with a $\gamma $-ray trapping parameter of $ t_{0\gamma}=122 $\,d. The central magnetar is estimated to have a magnetic field of $ B=4\times10^{15} $\,gauss and an initial spin period of $ 1.2 $\,ms. The magnetar properties needed to adequately reproduce the steep and luminous light curve are fairly extreme, but within theoretical limits \citep[see, e.g.,][]{2013MNRAS.431.1745G}. The estimated $ B $ value is higher than invoked for most models for SLSNe-I \citep{2017ApJ...850...55N}, and more consistent with magnetar powered long GRB models \citep{2007ApJ...659..561M}. \cite{2004ApJ...611..380T} suggested that a magnetar with such extreme parameters can spin down more rapidly than simple vacuum dipole spin-down, and that a rotational energy of up to $ \sim10^{52} $\,erg can be extracted during first $ \sim10 $\,s after the birth of the protoneutron star, which may also significantly affect the SN shock dynamics.

\subsection{Circumstellar interaction}
Finally, we consider ejecta-CSM interaction as an alternative powering source for the luminous light curve of \sn. One common example of ejecta-CSM interactions are SNe~IIn, which are generally characterised by relatively narrow emission lines ($ \rm FWHM\approx 10^2$--$10^3$\,\kms) in their spectra \citep{1990MNRAS.244..269S}. Depending on the CSM density and wind-velocity profile, shock interactions can make SNe substantially more luminous, with a wide range of \lc\ shapes. CSM interactions have been proposed for most of the LSNe-II to account for their high luminosities (e.g., SNe~2013fc \citealt{2016MNRAS.456..323K}; PTF10iam, \citealt{2016ApJ...819...35A}; ASASSN-15nx, \citealt{2018ApJ...862..107B}; and 2016gsd, \citealt{2020MNRAS.493.1761R}). 

In \sn, no relatively narrow lines are detected in the spectra after the first at +2.4\,d. Nor do we see any other signatures
proposed for CSM interaction scenarios contrived to hide narrow emission lines, like high-velocity \Hi\ absorption components \citep[see, e.g.,][]{2007AIPC..937..357C,2012MNRAS.422.1122I,2013MNRAS.433.1871B,2015ApJ...806..160B} or enhancement of the blue continuum \citep[e.g.,][]{2009MNRAS.400..866C,2012MNRAS.426.1905S,2018ApJ...862..107B}.
Using the flash-ionised lines in the first spectrum (see \S\ref{sec:fi_bf}), we estimated a mass-loss rate of $ \sim 2\times 10^{-4}\, \rm \msun\,yr^{-1} $, which is larger than for red supergiants \citep[$\dot{M}\lesssim 10^{-4}\,\msun\,\hbox{yr}^{-1}$; e.g.,][]{1988A&AS...72..259D, 2020MNRAS.492.5994B} but is much lower than required in interaction-powered SNe~IIn \citep[$ \dot{M}\gtrsim 10^{-3}\, \rm \msun\,yr^{-1} $; e.g.,][]{2014ARA&A..52..487S,2017hsn..book..403S}.
The luminosity available from CSM interactions is
\begin{multline}
L_{\rm CSM} \approx \dot{M} v_s^3 v_w^{-1}
= 2.1 \left[ { \dot{M} \over 10^{-4}\, {\rm M}_\odot\,\hbox{yr}^{-1}} \right]
\left[ { v_s \over 10^4\,\kms } \right]^3\\
\left[ { 30\,\kms \over v_w } \right]
\times 10^{42}\,\ergs,
\end{multline} 
which would only be sufficient to power the light curve
after $\sim 50$\,d.  However, with the relatively low
densities at this point, the shock would have difficulty
thermalising this energy to produce the observed optical
emission and the energy would more likely be radiated as 
X-rays.  Using the photospheric velocity ($\sim 6000$\,\kms)
rather than the H$\alpha$ velocity ($\sim 10,000$\,\kms) 
would reduce the shock luminosity by another factor of five.
Moreover, a light curve powered by interaction with a low-velocity wind (as inferred from flash-ionisation lines; see \S\ref{sec:fi_bf}) would have produced prominent narrow lines of $ \sim100\,\kms $ widths in the later spectra, which are clearly missing.

The early X-ray detections during +11\,d to +14\,d with a luminosity of $ \sim5\times10^{41}\,\ergs $ and the nondetections thereafter indicate even lower CSM densities if the shock luminosity emerges as X-rays. The detections are, however, substantially more luminous than seen in typical SNe~IIP/L \citep[$ \sim10^{38}$--$10^{39}\,\ergs $; e.g.,][]{2012MNRAS.419.1515D,2019ApJ...873L...3B}. This suggests that the progenitor of \sn\ had a relatively denser CSM than typical red supergiant progenitors of SNe~IIP/L.
In any case, both the flash-ionised spectrum and the X-ray measurements indicate that the density of the CSM is too low to drive an interaction-powered light curve.

\subsection{Inferred large \nickel\ mass}
As discussed in \S\ref{sec:model_rad} and \S\ref{sec:model_magnetar}, radioactive \nickel\ is a key component for either model to fit the light curve, especially during the tail phase. The radiative diffusion model requires a \nickel\ mass of $\sim 0.4 $\,\msun, while the magnetar spin-down model requires a slightly lower mass of $\sim 0.3 $\,\msun. Such masses are about an order of magnitude higher than found in typical SNe~IIP/L (the median is a few percent of $\msun$), and are also considerably higher than estimates for SNe~IIb \citep{CITE3,2017ApJ...841..127M,2019A&A...628A...7A}. 
Stripped ccSNe (SNe~Ib/c) tend to have higher \nickel\ masses than SNe~II \citep{CITE3, 2019A&A...628A...7A, 2020arXiv200407244S}. The \nickel\ mass estimates for \sn\ are within the range reported for SNe~Ib/c \citep{CITE3, 2019A&A...628A...7A} and are remarkably similar to estimates for SNe~Ic-BL, which have some of the highest estimated \nickel\ masses among ccSNe.

The nebular-phase spectrum of \sn\ shows strong lines of iron-group elements (as discussed in \S\ref{sec:spec_compare}) which are not seen in SNe~II, but are comparable to those in SNe~Ic-BL like SNe~1998bw and 2002ap which have \nickel\ mass estimates of $ \sim0.35\,\msun $ \citep{2001ApJ...550..991N} and $ \sim0.1\,\msun $ \citep{2007ApJ...670..592M} (respectively) based on detailed spectroscopic modeling. These strong lines also furnish an independent confirmation that \sn\ has produced a massive amount of \nickel\ in the explosion. However, detailed modeling is required to quantify this \nickel\ mass using nebular spectra.
\sn\ is a partially stripped-envelope SN, as inferred from the presence of helium and the nebular-phase spectra with weak hydrogen. The partially stripped envelope and the high expansion velocity lead to fast rarefication of the ejecta, consistent with the low $ \gamma $-ray trapping parameter $ t_{0\gamma} $ we estimated from light curve models.

The popular neutrino-driven explosion models are unable to produce $ M_{\rm Ni} $ higher than $ \sim0.2 $\,\msun\ \citep[e.g.,][]{2016ApJ...821...38S,2020ApJ...890...51E}. This poses a serious challenge for explaining many ccSNe, particularly stripped ccSNe with high $ M_{\rm Ni} $ estimates reported in the literature \citep[see e.g.,][]{2012ApJ...749L..28V,2019A&A...628A...7A}. Sometimes, the magnetar model is invoked to substantially reduce the high \nickel\ mass \citep[e.g.,][]{2016ApJ...831...41W}. However, this does not work for \sn, as the magnetar model only slightly reduced the required $ M_{\rm Ni} $ to 0.3\,\msun, which is still inconsistent with neutrino-driven explosion models. On the other hand, $ M_{\rm Ni} \gtrsim 0.2 $\,\msun\ may be synthesised by a strong magnetar itself as suggested by \cite{2015MNRAS.451..282S}, but our best-fit magnetar parameters for ASASSN-18am do not satisfy their constraint.  
Another possibility is the collapse-induced thermonuclear explosion models for ccSNe \citep{2015ApJ...811...97K, CITE3, CITE2}, which can produce such large \nickel\ masses and high kinetic energies (\citealt{CITE2}). Pair-instability SN models are also known to produce very large amounts of \nickel\ but have extended light curves peaking after several tens of days \citep{2011ApJ...734..102K}, incompatible with \sn.

\subsection{Fallback accretion power}
Fallback accretion onto a black hole following a neutrino-driven explosion of a massive progenitor may produce light curves with a wide range of luminosities. \citep[e.g.,][]{2010ApJ...723L..89U, 2019ApJ...880...21M}.
In such a model any \nickel\ produced in the shock is accreted 
onto the black hole remnant without contributing to radioactive heating, and acceretion power is the only source of energy powering the late-time light curves.
\cite{2019ApJ...880...21M} modeled the explosion of a $ 40 $\,\msun\ progenitor and produced a range of light curves depending on the accretion efficiency and the delay time of the fallback. These models could reproduce the light curves of normal-luminosity SNe~II (SN~1987A and SN~1999em), relatively luminous SNe~II (OGLE-2014-SN-073, \citealt{2017NatAs...1..713T}; SN~2009kf, \citealt{2010ApJ...717L..52B}), and also SLSN-II \cite[SN~2008es,][]{2009ApJ...690.1303M}. The fallback accretion powering mechanism is also a possibility for \sn. However, unlike the light curve of ASASSN-18am, these models show a long rise-to-peak time and slow decline rates. Moreover, as a consequence of the fallback, these models predict very low or no \nickel\ mass to be present in the ejecta, which contradicts the presence of strong nebular lines of iron-group elements suggesting a high \nickel\ mass yield. %

\section{Oxygen mass and nebular emission} \label{sec:oxygen_mass}
\begin{figure}
	\centering
	\includegraphics[width=0.91\linewidth]{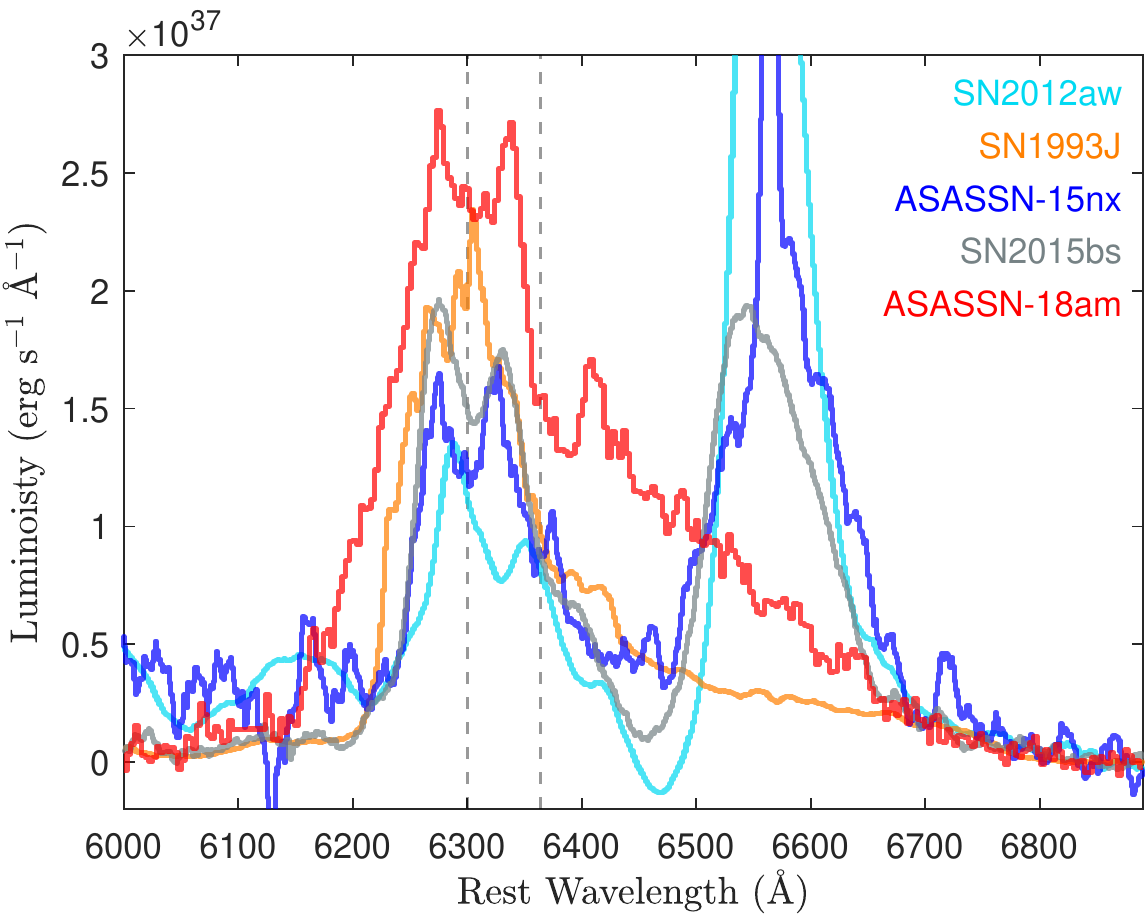}
	\caption{{The luminosity of the [\Oi] emission profile of \sn\ as compared to that of other SNe~II. For the comparison sample, the spectroscopic flux is dereddened and is recalibrated using available photometry. All of the spectra have the local continuum subtracted using a line-free region near 6850~\AA. The phases and references for the spectra are the same as in Fig.~\ref{fig:sp_comp_d211}.}}
	\label{fig:oi_compare}
\end{figure}

The strength of the [\Oi]~\ldld6300,~6364 emission line is directly related to the mass of oxygen produced in the core, which in turn depends on the ZAMS mass \citep[e.g.,][]{1995ApJS..101..181W,1996ApJ...460..408T}. As previously noted, the $ +211 $\,d nebular spectrum shows unusually strong [\Oi] emission as compared to other lines (e.g., \Feii, \Hi, \Hei, \Caii). In Figure~\ref{fig:oi_compare} the [\Oi] emission profile of \sn\ is compared in luminosity with that of other hydrogen-rich SNe. The [\Oi] luminosity of \sn\ is brighter than that of SN~II~2012aw, SN~IIb~1993J, LSN-II ASASSN-15nx, and SN~II~2015bs. SN~2015bs is claimed to have had one of the highest mass progenitors for a SN~II \citep[$ \sim21\,\msun $;][]{2018NatAs...2..574A}. SN~2012aw is an archetypal SN~IIP having direct identification of the progenitor from {\it HST} images with a mass of $ \sim18 $\,\msun\ \citep{2012ApJ...756..131V,2012ApJ...759L..13F}. This indicates that \sn\ has a high \Oi\ mass and consequently a massive progenitor when compared with most SNe~II.

{However, the temperature and line opacities must also be considered in order to accurately estimate the \Oi\ mass.} We can estimate the \Oi\ mass using the method described by \cite{2014MNRAS.439.3694J}. This requires an estimate of the \Oi\ temperature, which can be done using the line ratios of [\Oi]~\ld5577 and [\Oi]~\ldld6300,~6364, as both are collisionally excited but with different temperature dependencies. In our nebular spectrum [\Oi]~\ld5577 is marginally detected and partially blended with [\Feii]~\ld5528. Using a two-component Gaussian model, we deblend the lines to estimate an [\Oi]~\ld5577 luminosity of $ 0.9\times10^{38}\, \rm erg\,s^{-1}$. {Since the [\Oi]~\ld5577 detection is marginal and a higher [\Oi]~\ld5577 luminosity relative to [\Oi]~\ldld6300,~6364 leads to a lower estimate of the oxygen mass, we adopted an upper limit to the [\Oi]~\ld5577 luminosity of $ 1.5\times10^{38}\, \rm erg\,s^{-1}$, corresponding to the nominal value plus three times the uncertainty estimate, to produce a conservative (i.e., biased low) estimate of the oxygen mass.
}

\begin{figure}
\centering
\includegraphics[width=0.8\linewidth]{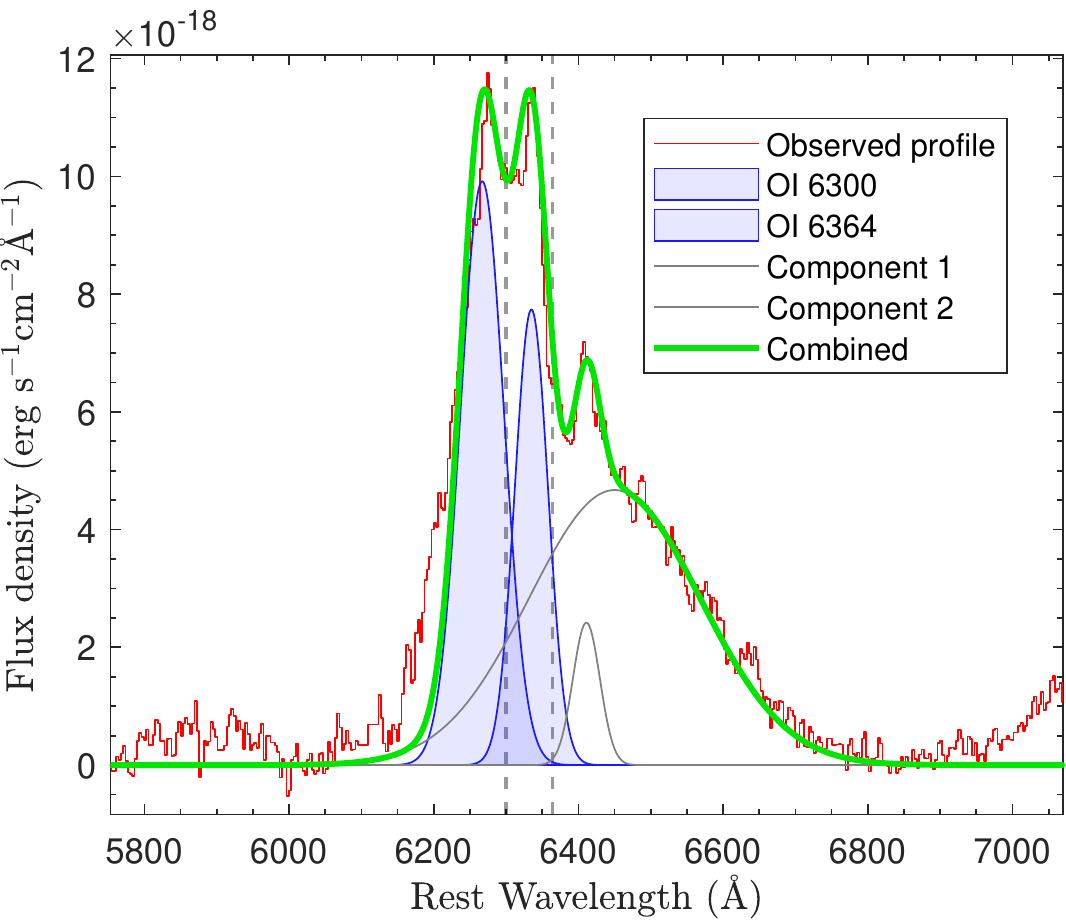}
\caption{Decomposition of the blended [\Oi]~\ldld6300,~6364 emission profile in the $ +211 $\,d nebular-phase spectrum. The spectrum pseudocontinuum has been subtracted, and a reddening correction was applied. The vertical dashed lines show the rest wavelengths of the [\Oi] doublet.}
\label{fig:oi_fit}
\end{figure}

\begin{figure}
	\centering
	\includegraphics[width=1\linewidth]{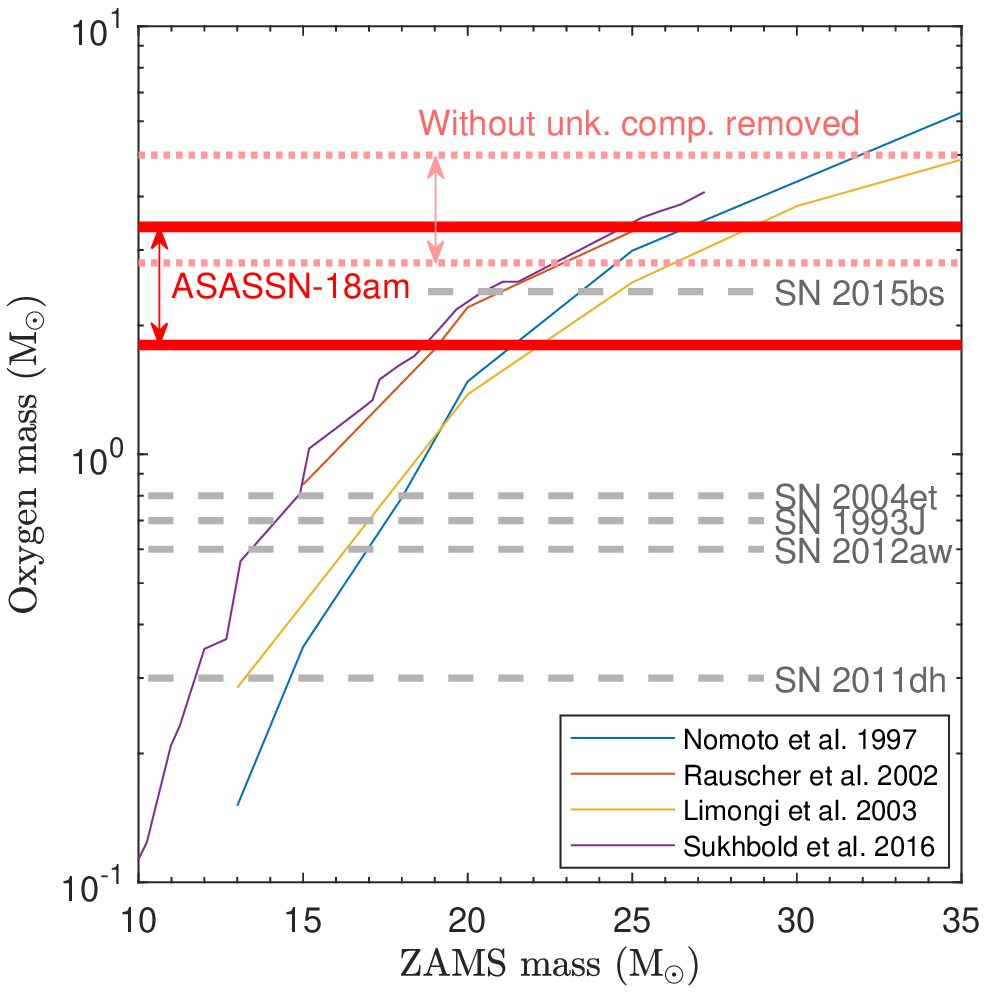}
\caption{{The final oxygen mass as a function of the progenitor ZAMS mass for the models of  \citet{1997NuPhA.616...79N}, \citet{2002ApJ...576..323R}, \citet{2003ApJ...592..404L}, and \citet{2016ApJ...821...38S}. The pair of red solid lines shows the range of \Oi\ mass estimated for \sn. The red dashed lines indicate the increase in the resulting \Oi\ mass if we assume there is no contamination of the line (i.e., unknown components not removed; see Fig.\ref{fig:oi_fit}). For comparison, the \Oi\ mass estimates for the normal SNe~II 2004et, 2012aw, 1993J, and 2011dh \citep{2012A&A...546A..28J,2014MNRAS.439.3694J,2015A&A...573A..12J} are denoted by grey dashed lines. We also estimated the \Oi\ mass for SN~2015bs to be $\sim 1.7$--3.1\,\msun\ using the spectrum from \citet{2018NatAs...2..574A}, and the implied ZAMS mass is consistent with their estimate.}}
	\label{fig:oi_models}
\end{figure}

Estimating the [\Oi]~\ldld6300,~6364 flux is nontrivial because of strong and broad emission (possibly a blend of \Hei, [\Nii], and \ha) on the red wing of the profile. As shown in Figure~\ref{fig:oi_fit}, we use four components to fit the full profile, where two of the components are for the 6300,~6364\,\AA\ doublet, and the other two (a narrow and a broad component) are to model the additional blended flux (see more discussion below). We measure a combined [\Oi] doublet luminosity of $ 2.8\times10^{39}\,\rm erg\,s^{-1}$. {In the optically thick limit, the [\Oi] $\lambda$6300/[\Oi] $\lambda$6340 line intensity ratio is $ \sim 1 $, whereas in the optically thin limit the ratio is $ \sim 3 $.} From our multicomponent fit we measure the 6300/6364 line ratio to be $ \sim1.2 $, which implies that the doublet emission is {partially transitioning to optical thinness}. Therefore, we follow the arguments by \cite{2014MNRAS.439.3694J} and adopt \cite{1957SvA.....1..678S} escape probabilities of  $ \beta_{6300,6364}\approx0.5 $ and $ \beta_{5577}/\beta_{6300,6364} \approx 1$--2. From this we estimate that the \Oi\ temperature is 3600--4000\,K and finally obtain an oxygen mass of $ M_{\rm O}=1.8$--3.4\,\msun. {This is significantly more massive than most normal SNe~II/IIb (e.g., SNe~2004et, 2012aw, 1993J, and 2011dh).
As shown in Figure~\ref{fig:oi_models},} the oxygen mass yield varies monotonically with the initial progenitor $ M_{\rm ZAMS} $. {Based on these scaling relations,} the estimated \Oi\ mass implies an initial mass of $ M_{\rm ZAMS}=19$--26\,\msun\ or $ M_{\rm ZAMS}=22 $\,\msun\ by assuming a mean $ \beta $ ratio of $ 1.5 $. 
Since most of the oxygen is produced during the hydrostatic burning phase, changes in the explosion physics should have little effect on \Oi\ to ZAMS mass-scaling relation. 
{Note that the dominant sources of uncertainties in the \Oi\ mass estimates are the measurements of [\Oi]~\ld5577 and [\Oi]~\ldld6300,~6364 line fluxes. However, both of these measurements were made such as to ensure that we obtain a conservative estimate of the resulting \Oi\ mass. If the extended red wing in the [\Oi]~\ldld6300,~6364 profile is a blend of multiple emission lines in that region, then we would expect an asymmetric and red-skewed profile for the unknown broad component (refer to Fig.\ref{fig:oi_fit}) implying a larger [\Oi] flux and an optically thick line (i.e., $ \beta_{6300,6364}<0.5 $), which will only result in a higher derived \Oi\ mass.}

We also examined the [\Oi]\ldld6300, 6364/[\Caii]\ldld7291, 7324 line intensity ratio. This ratio is sensitive to the core mass and hence to the initial progenitor mass, while being minimally dependent on temperature and density \citep{1989ApJ...343..323F,2004A&A...426..963E}. In SNe~IIP/L the line ratio is typically $ \lesssim0.7 $ \citep{2015A&A...579A..95K}. For \sn, we estimate that the [\Oi]/[\Caii] line ratio is $ \sim 2.3 $, also implying a significantly higher progenitor mass. To further compare the line ratio with that of various progenitor masses, we use the model spectra from \cite{2014MNRAS.439.3694J} for the mass range 12--25\,\msun\ and extract the [\Oi]/[\Caii] line ratios. After 250\,d past explosion, the ratios are found to be almost constant.
The ratio of $ \sim 2.0 $ for the highest $ 25\,\msun $ model is closest to (albeit slightly lower than) the value we find for \sn.
For the lower mass models, the line ratio monotonically decreases.

On modeling the [\Oi] emission in the +211\,d spectrum with multicomponent Gaussian profiles (see Fig.~\ref{fig:oi_fit}), we find that the doublet is blueshifted by $ \sim 1350 $\,\kms. A similar blueshift is also seen for the [\Caii] doublet. As mentioned above, for the decomposition of the [\Oi] profile we used two more components in addition to the [\Oi] doublet. One is to fit the weak narrow component at $ \sim6408\,$\AA\ (\mbox{Component-1} of Figure~\ref{fig:oi_fit}) and the other is to fit the excess flux on the blue wing (\mbox{Component-2}). We could not ascertain the exact origin of these two components, but we speculate these are likely blends of \Hei, [\Nii], and also possibly residual \ha. 
Another possibility is that \mbox{Component-1} (narrow) is a redshifted component of double-peaked [\Oi] emission. However, from our spectrum we cannot firmly determine any presence of double-peaked structure in the [\Oi] and [\Caii] profiles, which would otherwise imply a bipolar core geometry.
The observed blueshift can be a consequence of asymmetry in the inner ejecta, residual opacity in the core, or possibly the formation of dust \citep{2009MNRAS.397..677T}.
If \mbox{Component-1} is associated with [\Oi] due to bipolar geometry or if the blueshifted emission is a consequence of dust formation, then both of these scenarios would imply an even higher oxygen mass and consequently a higher ZAMS mass.

{A progenitor of $ M_{\rm ZAMS}\approx20$--25\,\msun\ will explode with a helium-core mass of $ M_{\rm He}\approx6$--8\,\msun\ \citep{2004A&A...425..649H}. Therefore, including the mass of the hydrogen envelope, the progenitor mass at the time of explosion will be $ >6$--8\,\msun. However, in the radiative diffusion model the estimated ejecta mass is $ \lesssim4 $\,\msun\ (see \S\ref{sec:model_rad}), and the implied mass of the compact remnant will be $ >2$--4\,\msun. Given the overall systematic uncertainties in mass estimates, the nature of the resulting remnant is uncertain, but a black hole may be favoured.}

\section{Summary} \label{sec:summary}
We presented discovery and follow-up observations of the luminous hydrogen-rich SN \sn. The light curve peaked at $M_V \approx -19.7 $\,mag, which is between that of normal ccSNe and SLSNe. The photospheric phase light curve exhibits a very steep decline of $ 6.0 $\,\maghundred, making it one of the fastest-declining SNe~II. 
The earliest spectrum at +2.4\,d shows flash-ionised features of \Hi\ and \Heii, after which the spectra became featureless with only a blue continuum until 18.2\,d. \sn\ is the first LSN-II having a spectrum sufficiently early to see the flash-ionisation features and from this we estimated that the star had a CSM wind of $ \dot{M} \approx 2\times10^{-4}\,\msun\,\hbox{yr}^{-1} $.
The early X-ray detections imply lower mass-loss rates but the X-rays may be partly thermalised at these phases. The later X-ray nondetections would seem to require lower CSM densities. 

By spectroscopic definition \sn\ is a Type~IIb SN because of persistent \Hei\ lines identified in its spectra.
The presence of helium and weak unresolved \ha\ at late phases suggests partially hydrogen-depleted ejecta. Other than these \Hei\ lines, \sn\ is both photometrically and spectroscopically different from generic SNe~IIb. Its overall energy budget is also significantly higher.
In SNe~IIb, the Balmer emission lines decay quickly after peak brightness and \Hei\ starts to dominate in the 6300--6900\,\AA\ region, but \sn\ has stronger Balmer emissions than SNe~IIb at coeval epochs. This seems to place \sn\ between generic SNe~IIP/L and SNe~IIb in terms of the hydrogen content of its ejecta. The expansion velocities measured for \ha, \hb, and \Feii\ in \sn\ are unusually high for a SN~II. The earliest \ha\ absorption-minimum velocity is  17,000\,\kms, which is never seen in normal SNe~II. The nebular-phase spectra showed very strong [\Oi]~\ldld6300, 6364 emission, suggesting a massive progenitor. Using the [\Oi] luminosity we estimate the \Oi\ core mass to be $ \sim1.7$--3.1\,\msun\ which corresponds to a progenitor mass of  $ \sim19$--24\,\msun. We also found a high [\Oi]/[\Caii] line ratio of $ \sim 2.3 $, exceeding by a factor of three that of typical SNe~IIP/L, which also suggests a massive progenitor. %

We investigated a range of possible powering mechanisms for \sn.
Both the radiative diffusion and magnetar spin-down model support a low-mass envelope with high kinetic energy, which is consistent with our spectroscopic observations. The radiative diffusion model would require a large \nickel\ mass of $ 0.4 $\,\msun\ and a high $ \gamma $-ray leakage rate to fit the light curve. The magnetar spin-down model requires slightly lower values for both the nickel mass ($ 0.3 $\,\msun) and $ \gamma $-ray leakage. The nebular-phase spectrum shows strong lines of iron-group elements, also indicating a high \nickel\ mass. Such strong line are similar to those in many SNe~Ic-BL but are never seen in typical SNe~II. However, the large \nickel\ masses estimated from both models are difficult or impossible to produce in a neutrino-driven explosion.
On the other hand, we could not find any evidence to support the CSM interaction scenario. The CSM density, whether derived from the flash-ionisation emission lines or the X-ray luminosities, is too low to produce the observed luminosity through CSM interaction.

\sn\ is the latest addition to the small number of luminous H-rich SNe. In this work we refer to them as LSNe-II. Other examples are PTF10iam \citep{2016ApJ...819...35A}, SN~2013fc \citep{2016MNRAS.456..323K}, ASASSN-15nx \citep{2018ApJ...862..107B}, and SN~2016gsd \citep{2020MNRAS.493.1761R}.
{Although these SNe (see Fig.~\ref{fig:mv_comp}) do not represent a statistically complete sample, however, it likely disproves the existence of a true gap between normal SNe and SLSNe as was previously thought \citep{2016ApJ...819...35A}.}
LSNe-II have peak absolute magnitudes of $ \sim-20 $\,mag, about 2--3\,mag more luminous than typical ccSNe. They all have relatively short rise-to-peak times of $ \sim15 $\,d, followed by a rapid decline in light curves. These SNe also have relatively weak \Hi\ line profiles compared to SNe~IIP/L. 
SNe with ejecta-CSM interaction (SNe~IIn) can be similarly luminous, but the absence of any obvious spectroscopic signature of interaction in LSNe-II distinguish them from the SN~IIn population, so their powering mechanism is an open question. 
In previous examples, CSM models with weak interaction signatures could still be invoked. However, for \sn\ we know that the system lacks a sufficiently dense CSM to account for its high luminosity.

We can update the rate estimate for LSNe-II-like events given by \cite{2018ApJ...862..107B} based on one event (ASASSN-15nx) and the ASAS-SN survey running time of 2.7\,yr. Now, with ASASSN-18am, we have two LSNe-II detected in the ASAS-SN survey. Counting through the end of year 2019, ASAS-SN has been running for 5.7\,yr. So, we can simply scale up the time-based rate estimate from \cite{2018ApJ...862..107B} by a factor of $ 2\times2.7/5.7\approx0.95 $. Therefore, the updated time-based rate would be $ r\approx27\rm\,Gpc^{-3}\,yr^{-1} $. An alternative rate estimate given by \cite{2018ApJ...862..107B} was to scale relative to the 499 SNe~Ia in ASAS-SN by the end of 2016. ASAS-SN has discovered 528 SNe~Ia through the end of 2017 \citep{2019MNRAS.484.1899H}, which scales to $ \sim700 $ SNe~Ia by the end of 2019. Thus, this second estimate is updated by a factor of $ 2\times449/700\approx1.28 $ to give $ r\approx22\rm\,Gpc^{-3}\,yr^{-1} $, consistent with the estimated based on survey time. This means that the LSN-II rate is comparable with the SLSN-I rate \citep[$ 91\rm\,Gpc^{-3}\,yr^{-1} $;][]{2017MNRAS.464.3568P}, which indicates a possible continuity in the luminosity function between normal ccSNe and SLSNe.

\section*{Acknowledgments}

An anonymous referee provided helpful suggestions for improving this paper.
We thank Boaz Katz and Amir Sharon for their assistance, and Dan Perley for reducing the Keck/LRIS +8.3\,d spectrum. K.Z.S. and C.S.K. are supported by NSF grants AST-1515927, AST-1814440, and AST-1908570. 
M.S. is supported by a project grant (8021-00170B) from the Independent Research Fund Denmark (IRFD) and by generous grants (13261 and 28021) from VILLUM FONDEN. B.J.S. is supported by NSF grants AST-1907570, AST-1908952, AST-1920392, and AST-1911074. Support for J.L.P. is provided in part by FONDECYT through grant 1191038 and by the Ministry of Economy, Development, and Tourism's Millennium Science Initiative through grant IC120009, awarded to The Millennium Institute of Astrophysics, MAS. T.A.T. is supported in part by NASA grant 80NSSC20K0531. K.A.A. is supported by the Danish National Research Foundation (DNRF132). Parts of this research were supported by the Australian Research Council Centre of Excellence for All Sky Astrophysics in 3 Dimensions (ASTRO 3D), through project number CE170100013. J.Z. is supported by the National Natural Science Foundation of China (grants 11803049, 41727803). A.V.F.'s supernova group is grateful for financial assistance from the Christopher R. Redlich Fund, the TABASGO Foundation, and the Miller Institute for Basic Research in Science (U.C. Berkeley).

ASAS-SN is supported by the Gordon and Betty Moore Foundation through grant GBMF5490 to the Ohio State
University, and by NSF grants AST-1515927 and AST-1908570. Development of ASAS-SN was supported by NSF grant AST-0908816, the Mt. Cuba Astronomical Foundation, the Center for Cosmology and AstroParticle Physics at the Ohio State University, the Chinese Academy of Sciences South America Center for Astronomy (CASSACA), the Villum Foundation, and George Skestos.
Some of the data presented herein were obtained at the W. M. Keck Observatory, which is operated as a scientific partnership among the California Institute of Technology, the University of California, and NASA; the Observatory was made possible by the generous financial support of the W. M. Keck Foundation. We thank the staffs at Lick and Keck Observatories for their assistance. Research at Lick Observatory is partially supported by a generous gift from Google. 

We acknowledge the support of the staff of the Xinglong 2.16\,m telescope. This work was partially supported by the Open Project Program of the CAS Key Laboratory of Optical Astronomy, National Astronomical Observatories, Chinese Academy of Sciences. We acknowledge the Telescope Access Program (TAP) funded by NAOC, CAS, and the Special Fund for Astronomy from the Ministry of Finance. This work is partly based on NUTS2 observations made with the NOT, operated by the NOT Scientific Association at the Observatorio del Roque de los Muchachos, La Palma, Spain, of IAC. ALFOSC is provided by IAA under a joint agreement with the University of Copenhagen and NOTSA. NUTS2 is funded in part by the Instrument Center for Danish Astrophysics (IDA). This work is partly based on observations made with the Gran Telescopio Canarias (GTC), installed at the Spanish Observatorio del Roque de los Muchachos of the Instituto de Astrofísica de Canarias, on the island of La Palma.

The LBT is an international collaboration among institutions in the United States, Italy, and Germany. LBT Corporation partners are The University of Arizona on behalf of the Arizona university system; Istituto Nazionale di Astrofisica, Italy; LBT Beteiligungsgesellschaft, Germany, representing the Max-Planck Society, the Astrophysical Institute Potsdam, and Heidelberg University; The Ohio State University; and The Research Corporation, on behalf of The University of Notre Dame, University of Minnesota, and University of Virginia.

\section*{Data availability}
The data underlying this article are available in the article and in its online supplementary material.

\bibliographystyle{mn2e}
\bibliography{ms}

\onecolumn
{\centering 
	\fontsize{2.4mm}{3.1mm}\selectfont
	\begin{longtable}
		{c c r c c c c c c l}
		\caption{Optical photometry of \sn\ in \textit{BVgriz} bands.}
		\label{tab:photsn}\\
		\hline
		UT Date     &JD      &Phase$^{a}$&$B$  &$V$  &$g$  &$r$  &$i$  &$z$  &Tel$^{b}$ \\
		(yyyy-mm-dd)&2458000+&(day)      &(mag)&(mag)&(mag)&(mag)&(mag)&(mag)& /Inst\\
		\hline
		\endfirsthead
		\multicolumn{3}{l}{ {\tablename\ \thetable\ - continued. }} \\
		\hline
		UT Date     &JD      &Phase$^{a}$&$B$  &$V$  &$g$  &$r$  &$i$  &$z$  &Tel$^{b}$ \\
		(yyyy-mm-dd)&2458000+&(day)      &(mag)&(mag)&(mag)&(mag)&(mag)&(mag)& /Inst\\
		\hline
		\endhead
		\hline
		\endfoot
		\hline
		\endlastfoot
2018-01-10.66   &  129.16   &    -1.4     &         ---         & non-detection <17.8 &         ---         &         ---         &         ---         &         ---         &           ASASSN \\
2018-01-11.66   &  130.16   &    -0.4     &         ---         & non-detection <17.6 &         ---         &         ---         &         ---         &         ---         &           ASASSN \\ \hline
2018-01-12.50   &  131.00   &     0.4     &         ---         &         ---         & 16.834 $\pm$ 0.086  &         ---         &         ---         &         ---         &           ASASSN \\
2018-01-13.65   &  132.15   &     1.5     &         ---         & 16.777 $\pm$ 0.126  &         ---         &         ---         &         ---         &         ---         &           ASASSN \\
2018-01-14.60   &  133.10   &     2.4     &         ---         & 16.832 $\pm$ 0.116  & 16.471 $\pm$ 0.064  &         ---         &         ---         &         ---         &           ASASSN \\
2018-01-15.56   &  134.06   &     3.4     &         ---         & 16.554 $\pm$ 0.096  & 16.363 $\pm$ 0.064  &         ---         &         ---         &         ---         &           ASASSN \\
2018-01-17.47   &  135.97   &     5.2     &         ---         & 16.428 $\pm$ 0.254  &         ---         & 16.325 $\pm$ 0.124  &         ---         &         ---         &         DN \\
2018-01-17.51   &  136.01   &     5.2     & 16.296 $\pm$ 0.042  & 16.405 $\pm$ 0.027  &         ---         & 16.452 $\pm$ 0.021  & 16.643 $\pm$ 0.019  &         ---         &               PO \\
2018-01-17.63   &  136.13   &     5.4     &         ---         & 16.363 $\pm$ 0.077  &         ---         &         ---         &         ---         &         ---         &           ASASSN \\
2018-01-18.47   &  136.97   &     6.2     & 16.244 $\pm$ 0.037  & 16.373 $\pm$ 0.051  &         ---         & 16.362 $\pm$ 0.041  & 16.502 $\pm$ 0.060  &         ---         &         DN \\
2018-01-18.55   &  137.05   &     6.3     &         ---         & 16.247 $\pm$ 0.042  & 16.227 $\pm$ 0.027  & 16.407 $\pm$ 0.026  & 16.508 $\pm$ 0.043  &         ---         &             Iowa \\
2018-01-19.49   &  137.99   &     7.2     & 16.103 $\pm$ 0.014  & 16.256 $\pm$ 0.017  &         ---         & 16.315 $\pm$ 0.018  & 16.445 $\pm$ 0.023  &         ---         &         LCOGT \\
2018-01-19.49   &  137.99   &     7.2     & 16.309 $\pm$ 0.072  & 16.308 $\pm$ 0.069  &         ---         & 16.380 $\pm$ 0.086  & 16.499 $\pm$ 0.069  &         ---         &         DN \\
2018-01-19.50   &  138.00   &     7.2     &         ---         &         ---         & 16.247 $\pm$ 0.050  &         ---         &         ---         &         ---         &           ASASSN \\
2018-01-19.55   &  138.05   &     7.2     &         ---         & 16.189 $\pm$ 0.045  & 16.167 $\pm$ 0.043  & 16.354 $\pm$ 0.037  & 16.445 $\pm$ 0.052  &         ---         &             Iowa \\
2018-01-20.46   &  138.96   &     8.1     & 16.229 $\pm$ 0.040  & 16.158 $\pm$ 0.053  &         ---         & 16.299 $\pm$ 0.043  & 16.500 $\pm$ 0.048  &         ---         &         DN \\
2018-01-20.51   &  139.01   &     8.2     & 16.158 $\pm$ 0.038  & 16.205 $\pm$ 0.026  &         ---         & 16.227 $\pm$ 0.018  & 16.352 $\pm$ 0.014  &         ---         &               PO \\
2018-01-20.56   &  139.06   &     8.2     &         ---         & 16.045 $\pm$ 0.032  & 16.117 $\pm$ 0.033  & 16.250 $\pm$ 0.025  & 16.307 $\pm$ 0.039  &         ---         &             Iowa \\
2018-01-20.56   &  139.06   &     8.2     &         ---         & 16.164 $\pm$ 0.069  & 16.153 $\pm$ 0.049  &         ---         &         ---         &         ---         &           ASASSN \\
2018-01-21.46   &  139.96   &     9.1     & 16.019 $\pm$ 0.017  & 16.133 $\pm$ 0.018  &         ---         & 16.195 $\pm$ 0.015  & 16.272 $\pm$ 0.024  &         ---         &         LCOGT \\
2018-01-22.55   &  141.05   &    10.1     & 16.090 $\pm$ 0.040  & 16.136 $\pm$ 0.029  &         ---         & 16.169 $\pm$ 0.050  & 16.286 $\pm$ 0.043  &         ---         &               PO \\
2018-01-22.56   &  141.06   &    10.1     &         ---         & 16.042 $\pm$ 0.036  & 16.094 $\pm$ 0.032  & 16.167 $\pm$ 0.026  & 16.321 $\pm$ 0.038  &         ---         &             Iowa \\
2018-01-23.55   &  142.05   &    11.1     &         ---         & 16.059 $\pm$ 0.042  & 16.064 $\pm$ 0.025  & 16.167 $\pm$ 0.024  & 16.204 $\pm$ 0.035  &         ---         &             Iowa \\
2018-01-24.55   &  143.05   &    12.1     & 16.083 $\pm$ 0.042  & 16.101 $\pm$ 0.028  &         ---         & 16.136 $\pm$ 0.031  & 16.215 $\pm$ 0.032  &         ---         &               PO \\
2018-01-25.46   &  143.96   &    13.0     & 16.185 $\pm$ 0.036  & 16.083 $\pm$ 0.037  &         ---         & 16.105 $\pm$ 0.033  & 16.249 $\pm$ 0.055  &         ---         &         DN \\
2018-01-25.55   &  144.05   &    13.0     &         ---         & 16.056 $\pm$ 0.042  & 16.082 $\pm$ 0.024  & 16.175 $\pm$ 0.022  & 16.159 $\pm$ 0.038  &         ---         &             Iowa \\
2018-01-26.55   &  145.05   &    14.0     & 16.132 $\pm$ 0.039  & 16.105 $\pm$ 0.027  &         ---         & 16.133 $\pm$ 0.029  & 16.192 $\pm$ 0.037  &         ---         &               PO \\
2018-01-26.55   &  145.05   &    14.0     &         ---         & 16.071 $\pm$ 0.047  & 16.087 $\pm$ 0.028  & 16.135 $\pm$ 0.028  &         ---         &         ---         &             Iowa \\
2018-01-27.51   &  146.01   &    14.9     & 16.163 $\pm$ 0.029  &         ---         & 16.115 $\pm$ 0.052  &         ---         &         ---         &         ---         &  ASASSN,DN \\
2018-01-28.46   &  146.96   &    15.9     &         ---         &         ---         & 16.244 $\pm$ 0.055  &         ---         &         ---         &         ---         &           ASASSN \\
2018-01-28.55   &  147.05   &    16.0     &         ---         & 16.121 $\pm$ 0.033  & 16.160 $\pm$ 0.029  & 16.146 $\pm$ 0.022  & 16.215 $\pm$ 0.028  &         ---         &             Iowa \\
2018-01-28.56   &  147.06   &    16.0     & 16.142 $\pm$ 0.044  & 16.163 $\pm$ 0.036  &         ---         & 16.158 $\pm$ 0.023  & 16.215 $\pm$ 0.013  &         ---         &               PO \\
2018-01-29.46   &  147.96   &    16.8     & 16.189 $\pm$ 0.021  & 16.216 $\pm$ 0.018  &         ---         & 16.224 $\pm$ 0.016  & 16.283 $\pm$ 0.021  &         ---         &         LCOGT \\
2018-01-29.54   &  148.04   &    16.9     & 16.263 $\pm$ 0.033  &         ---         &         ---         &         ---         &         ---         &         ---         &         DN \\
2018-01-29.55   &  148.05   &    16.9     &         ---         & 16.134 $\pm$ 0.046  & 16.135 $\pm$ 0.029  & 16.198 $\pm$ 0.031  & 16.146 $\pm$ 0.043  &         ---         &             Iowa \\
2018-01-29.61   &  148.11   &    17.0     &         ---         & 16.257 $\pm$ 0.116  &         ---         &         ---         &         ---         &         ---         &           ASASSN \\
2018-01-30.46   &  148.96   &    17.8     & 16.344 $\pm$ 0.034  & 16.182 $\pm$ 0.036  &         ---         & 16.252 $\pm$ 0.030  & 16.270 $\pm$ 0.032  &         ---         &         DN \\
2018-01-30.51   &  149.01   &    17.9     & 16.334 $\pm$ 0.040  & 16.203 $\pm$ 0.030  &         ---         & 16.171 $\pm$ 0.020  & 16.158 $\pm$ 0.018  &         ---         &               PO \\
2018-01-30.55   &  149.05   &    17.9     &         ---         & 16.132 $\pm$ 0.038  & 16.265 $\pm$ 0.029  & 16.252 $\pm$ 0.021  & 16.219 $\pm$ 0.028  &         ---         &             Iowa \\
2018-01-30.61   &  149.11   &    18.0     &         ---         & 16.145 $\pm$ 0.112  &         ---         &         ---         &         ---         &         ---         &           ASASSN \\
2018-01-31.43   &  149.93   &    18.7     & 16.438 $\pm$ 0.039  & 16.294 $\pm$ 0.032  &         ---         & 16.345 $\pm$ 0.034  & 16.348 $\pm$ 0.030  &         ---         &         DN \\
2018-01-31.55   &  150.05   &    18.9     &         ---         & 16.232 $\pm$ 0.045  & 16.291 $\pm$ 0.031  & 16.238 $\pm$ 0.028  & 16.253 $\pm$ 0.036  &         ---         &             Iowa \\
2018-01-31.56   &  150.06   &    18.9     & 16.402 $\pm$ 0.043  & 16.278 $\pm$ 0.029  &         ---         & 16.248 $\pm$ 0.020  & 16.304 $\pm$ 0.022  &         ---         &               PO \\
2018-02-02.45   &  151.95   &    20.7     & 16.566 $\pm$ 0.082  & 16.372 $\pm$ 0.104  &         ---         & 16.413 $\pm$ 0.093  & 16.373 $\pm$ 0.127  &         ---         &         DN \\
2018-02-02.51   &  152.01   &    20.8     & 16.506 $\pm$ 0.045  & 16.385 $\pm$ 0.030  &         ---         & 16.318 $\pm$ 0.021  & 16.234 $\pm$ 0.024  &         ---         &               PO \\
2018-02-04.50   &  154.00   &    22.7     & 16.657 $\pm$ 0.039  & 16.465 $\pm$ 0.051  &         ---         & 16.467 $\pm$ 0.041  & 16.407 $\pm$ 0.058  &         ---         &         DN \\
2018-02-04.51   &  154.01   &    22.7     & 16.645 $\pm$ 0.037  & 16.478 $\pm$ 0.027  &         ---         & 16.415 $\pm$ 0.016  & 16.428 $\pm$ 0.020  &         ---         &               PO \\
2018-02-05.47   &  154.97   &    23.6     &         ---         &         ---         & 16.528 $\pm$ 0.087  &         ---         &         ---         &         ---         &           ASASSN \\
2018-02-06.42   &  155.92   &    24.6     & 16.839 $\pm$ 0.046  & 16.589 $\pm$ 0.037  &         ---         & 16.584 $\pm$ 0.033  & 16.522 $\pm$ 0.052  &         ---         &         DN \\
2018-02-06.44   &  155.94   &    24.6     &         ---         &         ---         & 16.506 $\pm$ 0.084  &         ---         &         ---         &         ---         &           ASASSN \\
2018-02-06.51   &  156.01   &    24.6     & 16.751 $\pm$ 0.037  & 16.578 $\pm$ 0.027  &         ---         & 16.527 $\pm$ 0.023  & 16.497 $\pm$ 0.025  &         ---         &               PO \\
2018-02-08.48   &  157.98   &    26.6     &         ---         &         ---         & 16.730 $\pm$ 0.081  &         ---         &         ---         &         ---         &           ASASSN \\
2018-02-08.51   &  158.01   &    26.6     & 16.853 $\pm$ 0.038  & 16.660 $\pm$ 0.026  &         ---         &         ---         &         ---         &         ---         &               PO \\
2018-02-08.53   &  158.03   &    26.6     &         ---         & 16.574 $\pm$ 0.029  & 16.705 $\pm$ 0.018  & 16.617 $\pm$ 0.020  & 16.548 $\pm$ 0.029  &         ---         &             Iowa \\
2018-02-09.49   &  158.99   &    27.5     & 16.909 $\pm$ 0.046  & 16.690 $\pm$ 0.051  &         ---         & 16.603 $\pm$ 0.054  & 16.687 $\pm$ 0.035  &         ---         &         DN \\
2018-02-09.50   &  159.00   &    27.5     & 16.875 $\pm$ 0.019  & 16.733 $\pm$ 0.016  &         ---         & 16.685 $\pm$ 0.014  & 16.636 $\pm$ 0.018  &         ---         &         LCOGT \\
2018-02-09.50   &  159.00   &    27.5     &         ---         &         ---         & 16.691 $\pm$ 0.074  &         ---         &         ---         &         ---         &           ASASSN \\
2018-02-09.54   &  159.04   &    27.6     &         ---         & 16.635 $\pm$ 0.027  & 16.780 $\pm$ 0.019  & 16.662 $\pm$ 0.023  & 16.605 $\pm$ 0.033  &         ---         &             Iowa \\
2018-02-10.40   &  159.90   &    28.4     & 16.969 $\pm$ 0.054  & 16.774 $\pm$ 0.055  &         ---         & 16.666 $\pm$ 0.058  & 16.643 $\pm$ 0.038  &         ---         &         DN \\
2018-02-10.44   &  159.94   &    28.5     &         ---         &         ---         & 16.862 $\pm$ 0.074  &         ---         &         ---         &         ---         &           ASASSN \\
2018-02-10.49   &  159.99   &    28.5     & 17.008 $\pm$ 0.044  & 16.806 $\pm$ 0.029  &         ---         & 16.679 $\pm$ 0.036  & 16.669 $\pm$ 0.061  &         ---         &               PO \\
2018-02-10.55   &  160.05   &    28.6     &         ---         & 16.678 $\pm$ 0.076  &         ---         &         ---         &         ---         &         ---         &             Iowa \\
2018-02-11.55   &  161.05   &    29.5     &         ---         & 16.708 $\pm$ 0.027  &         ---         &         ---         &         ---         &         ---         &             Iowa \\
2018-02-12.40   &  161.90   &    30.4     & 17.204 $\pm$ 0.045  & 16.823 $\pm$ 0.029  &         ---         & 16.805 $\pm$ 0.023  & 16.805 $\pm$ 0.043  &         ---         &         DN \\
2018-02-12.54   &  162.04   &    30.5     &         ---         & 16.713 $\pm$ 0.042  & 16.918 $\pm$ 0.023  & 16.763 $\pm$ 0.020  & 16.663 $\pm$ 0.031  &         ---         &             Iowa \\
2018-02-13.39   &  162.89   &    31.3     & 17.223 $\pm$ 0.096  & 16.921 $\pm$ 0.055  &         ---         & 16.927 $\pm$ 0.044  & 16.867 $\pm$ 0.055  &         ---         &         DN \\
2018-02-13.43   &  162.93   &    31.4     & 17.222 $\pm$ 0.046  & 16.993 $\pm$ 0.036  &         ---         & 16.827 $\pm$ 0.071  & 16.796 $\pm$ 0.060  &         ---         &         LCOGT \\
2018-02-13.49   &  162.99   &    31.4     & 17.181 $\pm$ 0.039  & 16.935 $\pm$ 0.027  &         ---         & 16.804 $\pm$ 0.022  & 16.807 $\pm$ 0.022  &         ---         &               PO \\
2018-02-13.54   &  163.04   &    31.5     &         ---         & 16.845 $\pm$ 0.027  & 17.002 $\pm$ 0.021  & 16.823 $\pm$ 0.021  & 16.738 $\pm$ 0.035  &         ---         &             Iowa \\
2018-02-15.44   &  164.94   &    33.3     & 17.311 $\pm$ 0.034  & 17.060 $\pm$ 0.038  &         ---         & 16.962 $\pm$ 0.038  & 16.964 $\pm$ 0.058  &         ---         &         LCOGT \\
2018-02-17.52   &  167.02   &    35.3     & 17.460 $\pm$ 0.047  & 17.150 $\pm$ 0.046  &         ---         & 16.991 $\pm$ 0.100  & 17.043 $\pm$ 0.103  &         ---         &         LCOGT \\
2018-02-17.55   &  167.05   &    35.4     & 17.484 $\pm$ 0.044  & 17.162 $\pm$ 0.033  &         ---         & 17.010 $\pm$ 0.028  & 16.967 $\pm$ 0.042  &         ---         &               PO \\
2018-02-21.52   &  171.02   &    39.2     &         ---         & 17.381 $\pm$ 0.038  & 17.752 $\pm$ 0.071  & 17.328 $\pm$ 0.040  & 17.055 $\pm$ 0.132  &         ---         &             Iowa \\
2018-02-22.41   &  171.91   &    40.1     & 17.824 $\pm$ 0.033  & 17.458 $\pm$ 0.028  &         ---         & 17.224 $\pm$ 0.019  & 17.172 $\pm$ 0.030  &         ---         &         LCOGT \\
2018-02-22.54   &  172.04   &    40.2     &         ---         & 17.356 $\pm$ 0.036  & 17.642 $\pm$ 0.027  & 17.224 $\pm$ 0.028  & 17.138 $\pm$ 0.040  &         ---         &             Iowa \\
2018-02-23.37   &  172.87   &    41.0     &         ---         & 17.430 $\pm$ 0.088  &         ---         & 17.340 $\pm$ 0.084  & 17.254 $\pm$ 0.079  &         ---         &         DN \\
2018-02-23.49   &  172.99   &    41.1     & 18.007 $\pm$ 0.050  & 17.520 $\pm$ 0.028  &         ---         & 17.277 $\pm$ 0.022  & 17.255 $\pm$ 0.016  &         ---         &               PO \\
2018-02-25.39   &  174.89   &    43.0     & 18.161 $\pm$ 0.059  & 17.580 $\pm$ 0.036  &         ---         & 17.327 $\pm$ 0.028  & 17.279 $\pm$ 0.038  &         ---         &         LCOGT \\
2018-02-25.50   &  175.00   &    43.1     & 18.232 $\pm$ 0.113  & 17.511 $\pm$ 0.086  &         ---         & 17.407 $\pm$ 0.084  & 17.472 $\pm$ 0.117  &         ---         &         DN \\
2018-02-26.36   &  175.86   &    43.9     & 18.279 $\pm$ 0.086  & 17.647 $\pm$ 0.059  &         ---         & 17.470 $\pm$ 0.045  & 17.446 $\pm$ 0.084  &         ---         &         DN \\
2018-02-27.47   &  176.97   &    45.0     & 18.289 $\pm$ 0.073  & 17.819 $\pm$ 0.079  &         ---         & 17.607 $\pm$ 0.052  & 17.469 $\pm$ 0.062  &         ---         &         DN \\
2018-02-27.49   &  176.99   &    45.0     & 18.303 $\pm$ 0.043  & 17.775 $\pm$ 0.028  &         ---         & 17.469 $\pm$ 0.024  & 17.442 $\pm$ 0.016  &         ---         &               PO \\
2018-02-27.53   &  177.03   &    45.0     &         ---         & 17.557 $\pm$ 0.051  & 17.891 $\pm$ 0.039  & 17.351 $\pm$ 0.025  &         ---         &         ---         &             Iowa \\
2018-03-02.41   &  179.91   &    47.8     & 18.683 $\pm$ 0.181  & 17.990 $\pm$ 0.104  &         ---         & 17.724 $\pm$ 0.064  & 17.597 $\pm$ 0.091  &         ---         &         DN \\
2018-03-02.53   &  180.03   &    47.9     &         ---         & 18.034 $\pm$ 0.066  & 18.263 $\pm$ 0.067  & 17.662 $\pm$ 0.038  &         ---         &         ---         &             Iowa \\
2018-03-03.35   &  180.85   &    48.7     & 18.725 $\pm$ 0.151  & 18.036 $\pm$ 0.092  &         ---         & 17.784 $\pm$ 0.069  & 17.634 $\pm$ 0.093  &         ---         &         DN \\
2018-03-03.47   &  180.97   &    48.9     &         ---         & 18.000 $\pm$ 0.085  & 18.296 $\pm$ 0.054  & 17.745 $\pm$ 0.038  &         ---         &         ---         &             Iowa \\
2018-03-04.44   &  181.94   &    49.8     & 18.599 $\pm$ 0.177  & 18.156 $\pm$ 0.065  & 18.387 $\pm$ 0.037  & 17.757 $\pm$ 0.032  & 17.703 $\pm$ 0.162  &         ---         &    DN,Iowa \\
2018-03-05.47   &  182.97   &    50.8     &         ---         & 18.240 $\pm$ 0.255  &         ---         & 17.824 $\pm$ 0.176  &         ---         &         ---         &             Iowa \\
2018-03-05.49   &  182.99   &    50.8     &         ---         & 18.094 $\pm$ 0.053  &         ---         & 17.770 $\pm$ 0.033  & 17.725 $\pm$ 0.040  &         ---         &               PO \\
2018-03-06.35   &  183.85   &    51.6     & 18.872 $\pm$ 0.152  & 18.124 $\pm$ 0.126  &         ---         & 17.884 $\pm$ 0.148  & 17.806 $\pm$ 0.209  &         ---         &         DN \\
2018-03-06.47   &  183.97   &    51.8     &         ---         & 18.180 $\pm$ 0.116  & 18.646 $\pm$ 0.097  &         ---         &         ---         &         ---         &             Iowa \\
2018-03-07.49   &  184.99   &    52.8     & 18.971 $\pm$ 0.058  & 18.390 $\pm$ 0.034  &         ---         & 17.923 $\pm$ 0.025  &         ---         &         ---         &               PO \\
2018-03-07.49   &  184.99   &    52.8     & 18.948 $\pm$ 0.089  & 18.324 $\pm$ 0.049  &         ---         & 17.953 $\pm$ 0.064  & 17.793 $\pm$ 0.058  &         ---         &         DN \\
2018-03-07.53   &  185.03   &    52.8     &         ---         & 18.227 $\pm$ 0.055  & 18.577 $\pm$ 0.053  & 17.885 $\pm$ 0.037  &         ---         &         ---         &             Iowa \\
2018-03-10.33   &  187.83   &    55.5     & 19.309 $\pm$ 0.094  & 18.542 $\pm$ 0.036  &         ---         & 18.115 $\pm$ 0.028  & 17.975 $\pm$ 0.038  &         ---         &         DN \\
2018-03-12.52   &  190.02   &    57.6     &         ---         & 18.438 $\pm$ 0.090  & 18.786 $\pm$ 0.169  & 18.191 $\pm$ 0.090  &         ---         &         ---         &             Iowa \\
2018-03-13.42   &  190.92   &    58.5     & 19.347 $\pm$ 0.051  & 18.656 $\pm$ 0.033  & 18.928 $\pm$ 0.121  & 18.270 $\pm$ 0.048  & 18.141 $\pm$ 0.062  &         ---         &    DN,Iowa \\
2018-03-14.32   &  191.82   &    59.4     & 19.335 $\pm$ 0.106  & 18.599 $\pm$ 0.067  &         ---         & 18.314 $\pm$ 0.063  & 18.199 $\pm$ 0.063  &         ---         &         DN \\
2018-03-14.47   &  191.97   &    59.5     & 19.503 $\pm$ 0.076  & 18.685 $\pm$ 0.037  &         ---         & 18.241 $\pm$ 0.030  & 18.110 $\pm$ 0.037  &         ---         &               PO \\
2018-03-14.52   &  192.02   &    59.6     &         ---         & 18.756 $\pm$ 0.068  & 19.091 $\pm$ 0.076  & 18.179 $\pm$ 0.044  &         ---         &         ---         &             Iowa \\
2018-03-15.32   &  192.82   &    60.3     & 19.397 $\pm$ 0.085  & 18.676 $\pm$ 0.040  &         ---         & 18.350 $\pm$ 0.041  & 18.236 $\pm$ 0.042  &         ---         &         DN \\
2018-03-15.43   &  192.93   &    60.5     &         ---         & 18.727 $\pm$ 0.054  & 19.075 $\pm$ 0.055  & 18.236 $\pm$ 0.047  &         ---         &         ---         &             Iowa \\
2018-03-16.47   &  193.97   &    61.5     & 19.552 $\pm$ 0.052  & 18.779 $\pm$ 0.039  &         ---         & 18.327 $\pm$ 0.036  & 18.315 $\pm$ 0.037  &         ---         &               PO \\
2018-03-16.51   &  194.01   &    61.5     &         ---         & 18.548 $\pm$ 0.089  & 19.215 $\pm$ 0.104  & 18.257 $\pm$ 0.068  &         ---         &         ---         &             Iowa \\
2018-03-18.43   &  195.93   &    63.4     &         ---         & 18.835 $\pm$ 0.053  & 19.102 $\pm$ 0.031  & 18.295 $\pm$ 0.033  &         ---         &         ---         &             Iowa \\
2018-03-20.23   &  197.73   &    65.1     & 19.822 $\pm$ 0.064  & 19.042 $\pm$ 0.038  & 19.367 $\pm$ 0.044  & 18.547 $\pm$ 0.028  & 18.467 $\pm$ 0.022  & 18.301 $\pm$ 0.032  &               LT \\
2018-03-20.48   &  197.98   &    65.4     &         ---         & 18.944 $\pm$ 0.037  &         ---         &         ---         &         ---         &         ---         &               PO \\
2018-03-21.49   &  198.99   &    66.3     &         ---         & 18.938 $\pm$ 0.042  & 19.408 $\pm$ 0.030  & 18.480 $\pm$ 0.058  &         ---         &         ---         &             Iowa \\
2018-03-22.48   &  199.98   &    67.3     & 19.759 $\pm$ 0.069  &         ---         &         ---         &         ---         &         ---         &         ---         &               PO \\
2018-03-23.34   &  200.84   &    68.1     & 19.956 $\pm$ 0.102  & 19.052 $\pm$ 0.069  &         ---         & 18.600 $\pm$ 0.034  & 18.607 $\pm$ 0.046  &         ---         &         LCOGT \\
2018-03-24.48   &  201.98   &    69.2     &         ---         &         ---         &         ---         & 18.568 $\pm$ 0.032  &         ---         &         ---         &               PO \\
2018-03-24.48   &  201.98   &    69.2     &         ---         & 19.005 $\pm$ 0.055  & 19.355 $\pm$ 0.060  & 18.518 $\pm$ 0.054  &         ---         &         ---         &             Iowa \\
2018-03-26.32   &  203.82   &    71.0     & 20.127 $\pm$ 0.376  & 19.127 $\pm$ 0.238  &         ---         & 18.576 $\pm$ 0.108  & 18.748 $\pm$ 0.092  &         ---         &         LCOGT \\
2018-03-27.49   &  204.99   &    72.2     &         ---         & 19.062 $\pm$ 0.051  & 19.550 $\pm$ 0.030  & 18.547 $\pm$ 0.039  &         ---         &         ---         &             Iowa \\
2018-03-28.48   &  205.98   &    73.1     &         ---         & 19.163 $\pm$ 0.053  & 19.556 $\pm$ 0.045  & 18.695 $\pm$ 0.059  &         ---         &         ---         &             Iowa \\
2018-03-29.48   &  206.98   &    74.1     &         ---         &         ---         &         ---         &         ---         & 18.783 $\pm$ 0.045  &         ---         &               PO \\
2018-04-02.48   &  210.98   &    78.0     &         ---         & 19.389 $\pm$ 0.045  &         ---         &         ---         &         ---         &         ---         &               PO \\
2018-04-05.16   &  213.66   &    80.6     & 20.458 $\pm$ 0.088  & 19.534 $\pm$ 0.034  & 19.949 $\pm$ 0.054  & 19.011 $\pm$ 0.018  & 18.890 $\pm$ 0.034  & 18.707 $\pm$ 0.034  &               LT \\
2018-04-05.46   &  213.96   &    80.9     &         ---         & 19.472 $\pm$ 0.098  & 19.829 $\pm$ 0.062  & 18.878 $\pm$ 0.058  &         ---         &         ---         &             Iowa \\
2018-04-05.48   &  213.98   &    80.9     &         ---         &         ---         &         ---         & 18.925 $\pm$ 0.035  &         ---         &         ---         &               PO \\
2018-04-06.48   &  214.98   &    81.8     &         ---         &         ---         &         ---         &         ---         & 18.986 $\pm$ 0.052  &         ---         &               PO \\
2018-04-08.40   &  216.90   &    83.7     &         ---         & 19.593 $\pm$ 0.039  &         ---         &         ---         &         ---         &         ---         &               PO \\
2018-04-12.18   &  220.68   &    87.4     & 20.601 $\pm$ 0.256  & 19.670 $\pm$ 0.129  & 20.091 $\pm$ 0.100  & 19.167 $\pm$ 0.047  & 19.113 $\pm$ 0.044  & 18.965 $\pm$ 0.080  &               LT \\
2018-04-14.14   &  222.64   &    89.3     & 20.566 $\pm$ 0.048  & 19.734 $\pm$ 0.027  & 20.134 $\pm$ 0.034  & 19.181 $\pm$ 0.026  & 19.121 $\pm$ 0.030  & 18.890 $\pm$ 0.031  &               LT \\
2018-04-18.12   &  226.62   &    93.1     & 20.704 $\pm$ 0.044  & 19.890 $\pm$ 0.041  & 20.246 $\pm$ 0.037  & 19.327 $\pm$ 0.034  & 19.272 $\pm$ 0.050  & 19.002 $\pm$ 0.041  &               LT \\
2018-05-05.06   &  243.56   &   109.6     & 21.217 $\pm$ 0.130  & 20.368 $\pm$ 0.055  & 20.840 $\pm$ 0.062  & 19.859 $\pm$ 0.026  & 19.852 $\pm$ 0.036  & 19.520 $\pm$ 0.062  &               LT \\
2018-05-08.06   &  246.56   &   112.5     & 21.389 $\pm$ 0.073  & 20.550 $\pm$ 0.049  & 20.908 $\pm$ 0.044  & 19.981 $\pm$ 0.035  & 19.943 $\pm$ 0.063  & 19.643 $\pm$ 0.083  &               LT \\
2018-05-18.02   &  256.52   &   122.1     & 21.422 $\pm$ 0.070  & 20.799 $\pm$ 0.052  & 21.159 $\pm$ 0.046  & 20.257 $\pm$ 0.046  & 20.376 $\pm$ 0.045  & 19.849 $\pm$ 0.060  &               LT \\
2018-05-23.00   &  261.50   &   127.0     & 21.685 $\pm$ 0.176  & 20.912 $\pm$ 0.088  & 21.218 $\pm$ 0.085  & 20.428 $\pm$ 0.040  & 20.434 $\pm$ 0.064  & 20.102 $\pm$ 0.098  &               LT \\
2018-06-02.97   &  272.47   &   137.6     & 21.842 $\pm$ 0.084  & 21.223 $\pm$ 0.051  & 21.478 $\pm$ 0.056  & 20.617 $\pm$ 0.033  & 20.773 $\pm$ 0.050  & 20.398 $\pm$ 0.071  &               LT \\
2018-06-08.02   &  277.52   &   142.5     &         ---         &         ---         &         ---         & 20.723 $\pm$ 0.037  &         ---         &         ---         &               LT \\
2018-06-21.07   &  290.57   &   155.2     & 22.137 $\pm$ 0.127  & 21.599 $\pm$ 0.089  & 21.858 $\pm$ 0.069  & 21.011 $\pm$ 0.060  & 21.211 $\pm$ 0.175  & 20.943 $\pm$ 0.158  &               LT \\
2018-06-24.04   &  293.54   &   158.0     &         ---         &         ---         &         ---         & 21.111 $\pm$ 0.084  &         ---         &         ---         &               LT \\
2018-07-03.04   &  302.54   &   166.8     &         ---         &         ---         &         ---         & 21.317 $\pm$ 0.102  &         ---         &         ---         &               LT \\
2018-07-14.96   &  314.46   &   178.3     &         ---         & 22.122 $\pm$ 0.121  &         ---         & 21.528 $\pm$ 0.079  &         ---         &         ---         &               LT \\
2018-08-17.89   &  348.39   &   211.2     & 23.300 $\pm$ 1.000  & 22.900 $\pm$ 0.500  & 22.700 $\pm$ 0.800  & 22.180 $\pm$ 0.400  & 22.332 $\pm$ 0.500  &         ---         &             SPEC \\
\hline
		
	\end{longtable}
	\addtocounter{table}{-1}
	\begin{longtable}
		{c c r c c c c l}
		\caption*{NUV photometry.}\\
		\hline
		UT Date     &JD      &Phase$^{a}$&$uvw2$&$uvm2$&$uvw1$&$uvu$&Tel$^{b}$ \\
		(yyyy-mm-dd)&2458000+&(day)      &(mag) &(mag) &(mag) &(mag)& / Inst\\
		\hline
		\endfirsthead
		\multicolumn{3}{l}{ {\tablename\ \thetable\ - continued. }} \\
		\hline
		UT Date     &JD      &Phase$^{a}$&$uvw2$&$uvm2$&$uvw1$&$uvu$&Tel$^{b}$ \\
		(yyyy-mm-dd)&2458000+&(day)      &(mag) &(mag) &(mag) &(mag)& / Inst\\
		\hline
		\endhead
		\hline
		\endfoot
		\hline
		\endlastfoot
2018-01-21.53  &   140.03   &    9.1    &  15.031 $\pm$ 0.042 &  14.811 $\pm$ 0.042 &  14.870 $\pm$ 0.044 &  14.974 $\pm$ 0.041 &   UVOT  \\
2018-01-23.86  &   142.36   &   11.4    &  15.112 $\pm$ 0.044 &  14.930 $\pm$ 0.043 &  14.853 $\pm$ 0.045 &  15.074 $\pm$ 0.044 &   UVOT  \\
2018-01-24.91  &   143.41   &   12.4    &  15.259 $\pm$ 0.041 &  15.023 $\pm$ 0.042 &  14.938 $\pm$ 0.039 &         ---         &   UVOT  \\
2018-01-26.05  &   144.55   &   13.5    &  15.509 $\pm$ 0.041 &  15.314 $\pm$ 0.043 &  15.151 $\pm$ 0.040 &         ---         &   UVOT  \\
2018-01-31.02  &   149.52   &   18.4    &  16.319 $\pm$ 0.053 &  16.025 $\pm$ 0.050 &  15.848 $\pm$ 0.052 &  15.511 $\pm$ 0.045 &   UVOT  \\
2018-02-02.16  &   151.66   &   20.4    &  16.662 $\pm$ 0.060 &  16.434 $\pm$ 0.055 &  16.097 $\pm$ 0.057 &  15.729 $\pm$ 0.051 &   UVOT  \\
2018-02-04.61  &   154.11   &   22.8    &  16.998 $\pm$ 0.085 &  16.831 $\pm$ 0.092 &  16.596 $\pm$ 0.084 &  15.955 $\pm$ 0.065 &   UVOT  \\
2018-02-06.01  &   155.51   &   24.2    &  17.639 $\pm$ 0.119 &  17.222 $\pm$ 0.096 &  16.778 $\pm$ 0.074 &  16.139 $\pm$ 0.060 &   UVOT  \\
2018-02-09.80  &   159.30   &   27.8    &  17.951 $\pm$ 0.130 &         ---         &  17.283 $\pm$ 0.091 &  16.596 $\pm$ 0.068 &   UVOT  \\
2018-02-12.71  &   162.21   &   30.7    &  18.546 $\pm$ 0.164 &  18.670 $\pm$ 0.180 &  17.873 $\pm$ 0.136 &  16.856 $\pm$ 0.080 &   UVOT  \\
2018-02-18.04  &   167.54   &   35.8    &  19.175 $\pm$ 0.241 &  20.036 $\pm$ 0.452 &  18.630 $\pm$ 0.215 &  17.477 $\pm$ 0.109 &   UVOT  \\
2018-02-23.75  &   173.25   &   41.4    &  20.774 $\pm$ 0.843 &         ---         &  18.894 $\pm$ 0.265 &  18.216 $\pm$ 0.187 &   UVOT  \\
2018-02-26.53  &   176.03   &   44.1    &  20.268 $\pm$ 0.533 &  20.232 $\pm$ 0.561 &  19.360 $\pm$ 0.370 &  18.716 $\pm$ 0.245 &   UVOT  \\
2018-02-28.52  &   178.02   &   46.0    &  21.059 $\pm$ 0.735 &         ---         &         ---         &  18.689 $\pm$ 0.180 &   UVOT  \\
2018-03-05.43  &   182.93   &   50.8    &         ---         &         ---         &  20.432 $\pm$ 0.696 &  19.254 $\pm$ 0.317 &   UVOT  \\
2018-03-06.49  &   183.99   &   51.8    &         ---         &         ---         &         ---         &  18.999 $\pm$ 0.294 &   UVOT  \\
\hline
		
	\end{longtable}
	\addtocounter{table}{-1}
	\begin{longtable}
		{c c r c c c l}
		\caption*{NIR photometry.}\\
		\hline
		UT Date     &JD      &Phase$^{a}$&$J$&$H$&$K$&Tel$^{b}$ \\
		(yyyy-mm-dd)&2458000+&(day)      &(mag) &(mag) &(mag) & /Inst\\
		\hline
		\endfirsthead
		\multicolumn{3}{l}{ {\tablename\ \thetable\ - continued. }} \\
		\hline
		UT Date     &JD      &Phase$^{a}$&$J$&$H$&$K$&Tel$^{b}$ \\
		(yyyy-mm-dd)&2458000+&(day)      &(mag) &(mag) &(mag) & /Inst\\
		\hline
		\endhead
		\hline
		\endfoot
		\hline
		\endlastfoot
2018-03-12.19  &   189.69   &    57.3   &   17.284 $\pm$ 0.047 &  17.012 $\pm$ 0.100  & 16.706 $\pm$ 0.163  &    NC \\
2018-03-30.57  &   208.07   &    75.1   &   18.019 $\pm$ 0.061 &  17.695 $\pm$ 0.042  &       ---           & UKIRT \\
2018-05-16.43  &   254.93   &   120.6   &   19.438 $\pm$ 0.081 &  18.448 $\pm$ 0.056  &       ---           & UKIRT \\
2018-06-03.60  &   273.10   &   138.2   &   19.942 $\pm$ 0.138 &  18.915 $\pm$ 0.096  &       ---           & UKIRT \\
2018-06-04.28  &   273.78   &   138.9   &         ---          &  18.861 $\pm$ 0.096  &       ---           & UKIRT \\
2018-06-28.44  &   297.94   &   162.3   &   20.515 $\pm$ 0.196 &  19.307 $\pm$ 0.112  &       ---           & UKIRT \\
\hline
		
	\end{longtable}

	\begin{flushleft}
			$^{a}$Rest frame days with reference to the explosion epoch \EpEpoch.\\
$^{b}$ The abbreviations of telescope/instrument used are as follows: ASASSN - ASAS-SN quadruple 14-cm telescopes; LCOGT - Las Cumbres Observatory 1 m telescope
	network; LT - 2m Liverpool Telescope; DN - 0.5m DEMONEXT telescope; PO - 0.6m telescopes of Post observatory; Iowa - 0.5m Iowa Robotic Telescope; SPEC - synthetic photometry using GTC spectrum;
	NC - NotCAM NIR imager mounted on 2.6m NOT; 
	UKIRT - WFCAM NIR imager mounted on 3.8m UKIRT; 
	UVOT - \swift\ Ultraviolet Optical Telescope.\\
	Data observed within 5\,hr are represented under a single-epoch observation.
	\end{flushleft}
}

\newpage

\begin{table}
\centering
  \caption{Summary of spectroscopic observations of \sn.}
  \label{tab:speclog}
  \begin{tabular}{lc r l}
    \hline
    UT Date       &JD $-$      &Phase$^{a}$&Telescope \\
                &2,458,000 & (day)     &/ Instrument                \\ \hline
2018-01-14.54 & 133.04 & 2.4   & FLWO/FAST \\
2018-01-20.27 & 138.77 & 7.9   & NOT/ALFOSC  \\
2018-01-20.66 & 139.16 & 8.3   & Keck/LRIS \\
2018-01-22.22 & 140.72 & 9.8   & NOT/ALFOSC  \\
2018-01-24.48 & 142.98 & 12.0  & Palomar/DBSP \\
2018-01-26.87 & 145.37 & 14.3  & Xinglong/BFOSC  \\
2018-01-30.87 & 149.37 & 18.2  & Xinglong/BFOSC  \\
2018-01-31.49 & 149.99 & 18.8  & MDM/OSMOS  \\
2018-02-01.52 & 151.02 & 19.8  & MDM/OSMOS  \\
2018-02-02.49 & 151.99 & 20.7  & MDM/OSMOS  \\
2018-02-09.54 & 159.04 & 27.6  & Shane/Kast  \\
2018-02-13.24 & 162.74 & 31.2  & NOT/ALFOSC  \\
2018-03-04.47 & 181.97 & 49.8  & MDM/OSMOS  \\
2018-03-23.26 & 200.76 & 68.0  & NOT/ALFOSC  \\
2018-04-05.15 & 213.65 & 80.6  & NOT/ALFOSC  \\
2018-05-18.35 & 256.85 & 122.5 & LBT/MODS  \\
2018-08-17.89 & 348.39 & 211.2 & GTC/OSRIS  \\
    \hline
  \end{tabular}
\begin{flushleft}
  $^{a}$Rest-frame days with reference to the explosion epoch \EpEpoch.\\
  See \S\ref{sec:data} for telescope and instrument details.
\end{flushleft}
\end{table}

\begin{table*}
	\centering
	\caption{X-ray detections and upper limits}
	\label{tab:xray_det}
	\begin{tabular}{ccrccl}
		\hline
		UT Date      & JD       & Phase$^a$ & Flux$^b$                     & Luminosity         & Tel   \\
		(yyyy-mm-dd) & 2458000+ & (day)     & ($10^{-13}~\rm ergs~s^{-1}cm^{-1}$) & ($10^{41}~\ergs$) &       \\ \hline
		2018-01-20   & 140.03   & 9.1       & $<1.9$                       & $<4.5$            & Swift \\
		2018-01-22   & 142.36   & 11.4      & $2.78\pm1.18$                & $6.5\pm2.8$       & Swift \\
		2018-01-23   & 143.41   & 12.4      & $1.82\pm0.93$                & $4.3\pm2.2$       & Swift \\
		2018-01-25   & 144.55   & 13.5      & $2.00\pm0.92$                & $4.7\pm2.2$       & Swift \\
		2018-01-30   & 149.52   & 18.4      & $<1.8$                       & $<4.1$            & Swift \\
		2018-02-01   & 151.66   & 20.4      & $<1.6$                       & $<3.7$            & Swift \\
		2018-03-14   & 192.19   & 59.7      & $<0.16$                      & $<0.38$           & Chandra \\ \hline
	\end{tabular}
	\begin{flushleft}
		$^{a}$ Rest frame days with reference to the explosion epoch \EpEpoch.\\
		$^b$ Fluxes in the 0.3-10 keV band. The errors and upper-limits are $ 1\sigma $ values.
	\end{flushleft}
\end{table*}

\appendix

\label{lastpage}

\onecolumn
\noindent  
$^{1}$Department of Astronomy, The Ohio State University, 140 W. 18th Avenue, Columbus, OH 43210, USA.\\
$^{2}$Center for Cosmology and AstroParticle Physics (CCAPP), The Ohio State University, 191 W. Woodruff Avenue, Columbus, OH 43210, USA.\\
$^{3}$Kavli Institute for Astronomy and Astrophysics, Peking University, Yi He Yuan Road 5, Hai Dian District, Beijing 100871, China.\\
$^{4}$Department of Physics and Astronomy, Aarhus University, Ny Munkegade 120, DK-8000 Aarhus C, Denmark \\
$^{5}$Department of Physics, Florida State University, Tallahassee, FL 32306, USA\\
$^{6}$INAF-Osservatorio Astronomico di Padova, Vicolo dell'Osservatorio 5, I-35122 Padova, Italy\\
$^{7}$Harvard-Smithsonian Center for Astrophysics, 60 Garden St., Cambridge, MA 02138, USA\\
$^{8}$Department of Astronomy, University of California, Berkeley, CA 94720-3411, USA\\
$^{9}$Miller Senior Fellow, Miller Institute for Basic Research in Science, University of California, Berkeley, CA 94720, USA.\\
$^{10}$N\'ucleo de Astronom\'ia de la Facultad de Ingenier\'ia y Ciencias, Universidad Diego Portales, Av. Ej \'ercito 441, Santiago, Chile\\
$^{11}$Millennium Institute of Astrophysics, Santiago, Chile.\\
$^{12}$Tuorla Observatory, Department of Physics and Astronomy, University of Turku, FI-20014 Turku, Finland\\
$^{13}$CAS Key Laboratory of Optical Astronomy, National Astronomical Observatories, Chinese Academy of Sciences, Beijing 100101, China\\
$^{14}$School of Physics, The University of Melbourne, Parkville, VIC 3010, Australia\\
$^{15}$ARC Centre of Excellence for All Sky Astrophysics in 3 Dimensions (ASTRO 3D)\\
$^{16}$DARK, Niels Bohr Institute, University of Copenhagen, Lyngbyvej 2, 2100 Copenhagen, Denmark\\
$^{17}$Department of Astronomy and Astrophysics, University of California, Santa Cruz, CA 95064, USA\\
$^{18}$Department of Astronomy and Astrophysics, University of California, Santa Cruz, California, 95064, USA\\
$^{19}$Dipartimento di Fisica e Astronomia `G. Galilei', Universit\`a di Padova, Vicolo dell'Osservatorio 3, I-35122 Padova, Italy\\
$^{20}$Dept. of Physics, Earth Science, and Space System Engineering, Morehead State Univ., 235 Martindale Dr., Morehead, KY 40351, USA\\
$^{21}$Carnegie Observatories, 813 Santa Barbara Street, Pasadena, CA 91101, USA\\
$^{22}$Department of Astronomy and The Oskar Klein Centre, AlbaNova University Center, Stockholm University, SE-10691 Stockholm, Sweden\\
$^{23}$Department of Physics and Astronomy, University of Iowa, Iowa City, IA 52242\\
$^{24}$Trinity University, Department of Physics \& Astronomy, One Trinity Place, San Antonio, TX 78212\\
$^{25}$Post Observatory, Lexington, MA 02421\\
$^{26}$Department of Physics and Astronomy, University of California, Riverside, CA 92507, USA\\
$^{27}$Institute for Astronomy, University of Hawaii, 2680 Woodlawn Drive, Honolulu, HI 96822, USA\\

\end{document}